\documentclass{aa}

\usepackage{graphicx}
\usepackage{multirow}
\usepackage{multicol}
\usepackage{txfonts }
\usepackage{natbib  }
\usepackage{hyperref}
\usepackage{breakurl}
\usepackage{subfig  }
\usepackage{xcolor  }
\usepackage{soul    }
\usepackage{float   }

\newcommand{\Jyb}{\textrm{Jy beam}$^{-1}$}

\begin{document}

\title{Thermal radio absorption as a tracer of the interaction of SNRs with their environments}

\author{G.    Castelletti\inst{1}
   \and L.    Supan\inst{1}
   \and W. M. Peters\inst{2}
   \and N. E. Kassim\inst{2}}

\institute{Instituto de Astronom\'ia y F\'isica del Espacio (IAFE, CONICET $-$ UBA) CC 67, Suc. 28, 1428 Buenos Aires, Argentina\\
\email{gcastell@iafe.uba.ar}
\and Code 7213, Remote Sensing Division, U. S. Naval Research Laboratory, 4555 Overlook Ave. SW, Washington, DC 20375, USA}

\date{Received 25 June 2021 / Accepted 14 August 2021}

\abstract
{
We present new images and continuum spectral analysis for 14 resolved Galactic SNRs selected from the 74~MHz Very Large Array Low-Frequency Sky Survey Redux (VLSSr). We combine new integrated measurements from the VLSSr with, when available, flux densities extracted from the Galactic and Extragalactic All-Sky Murchison Widefield Array Survey (GLEAM) and measurements from the literature to generate improved integrated continuum  spectra sampled from $\sim$15~MHz to $\sim$217~GHz. 
We present the VLSSr images.  When possible we combine them with publicly available images at 1.4~GHz, to analyse the resolved morphology and spectral index distribution across each SNR.  We interpret the results and look for evidence of thermal absorption caused by ionised gas either proximate to the SNR itself, or along its line of sight. Three of the SNRs, G4.5+6.8 (Kepler), G28.6$-$0.1, and G120.1+1.4 (Tycho), have integrated spectra which can be adequately fit with simple power laws. The resolved spectral index map for Tycho confirms internal absorption which was previously detected by the Low Frequency Array (LOFAR: \citealt{arias+19}), but it is insufficient to affect the fit to the integrated spectrum. Two of the SNRs are pulsar wind nebulae, G21.5$-$0.9 and G130.7+3.1 (3C~58).  For those we identify high-frequency spectral breaks at 38 and 12~GHz, respectively. For the integrated spectra of the remaining nine SNRs, a low frequency spectral turnover is necessary to adequately fit the data.  In all cases we are able to explain the turnover by extrinsic thermal absorption. For G18.8+0.3 (Kes~67), G21.8$-$0.6 (Kes~69), G29.7$-$0.3 (Kes~75), and G41.1$-$0.3 (3C~397), we attribute the absorption to ionised gas along the line of sight, possibly from extended HII region envelopes (EHEs). For G23.3$-$0.3 (W41) the absorption can be attributed to HII regions located in its immediate proximity. Thermal absorption from interactions at the ionised interface between SNR forward shocks and the surrounding medium were previously identified by \citet{bro05-391} as responsible for the low frequency turnover in SNR G31.9+0.0 (3C~391); our integrated spectrum is consistent with the previous results.  We present evidence for the same phenomenon in three additional SNRs G27.4+0.0 (Kes~73), G39.2$-$0.3 (3C~396), and G43.3$-$0.2 (W49B), and derive constraints on the physical properties of the interaction. This result indicates that interactions between SNRs and their environs should be readily detectable through thermal absorption by future low frequency observations of SNRs with improved sensitivity and resolution.  
}

\keywords{Radio continuum: general - ISM: supernova remnants}

\titlerunning{Interacting SNRs as traced by low radio frequency observations}
\maketitle

\section{Introduction}
\label{introduction}

Supernova remnants (SNRs) are the most prominent extended sources of non-thermal emission in the Galaxy and impact Galactic astrophysics in several important ways. They and their supernovae (SNe) progenitors play a key role in stellar evolution by marking the death of massive stars, redistributing the atomic elements produced within them, and stimulating the birth of new stars through their interaction with molecular clouds. SNR shocks significantly impact the dynamics and evolution of the interstellar medium (ISM) \citep{bre+17}, and play a major role in shaping its long lived structure \citep{mckee1977}. As the presumed primary acceleration sites for Galactic cosmic rays (CRs) through diffusive shock acceleration \citep{cap10,gabici+19,kac19}, they seed the ISM  with at least $\sim$1/3 of its energy density. A complete census of Galactic SNRs, an elusive goal long prevented by observational selection effects \citep{green1991}, would offer powerful constraints on the Galactic SNe rate. 

At radio wavelengths, SNRs are spatially extended and typically prolific non-thermal emitters over a broad range of frequencies.  Accurate radio spectra for SNRs can be used to trace interactions with the ISM, and to test predictions from diffusive shock acceleration and SNR evolution theories.  For largely technical reasons (\citealt{kassim07}, and references therein), low frequency ($\nu < 100$~MHz) radio observations of SNRs have historically been limited by extremely poor angular resolution and sensitivity, typically an order of magnitude or more worse than what is achieved at higher (GHz) frequencies.  The scientific impact for SNRs has been significant, limiting a number of unique and important studies that critically rely on precise low frequency measurements. These include: 
{\it (i)\,} Deviations from power law spectra for SNRs predicted by theory, which can only be constrained by measurements encompassing a very large range of frequencies (\citealt{reynolds92}).  Because the deviations are subtle, a high degree of accuracy is needed. For most SNRs these have been unavailable until only recently \citep{urosevic2014,arias+19-VRO}. 
{\it (ii)\,}  Thermal absorption in and around SNRs, which is uniquely measured at low frequencies ($\nu < 100$~MHz).  It may arise intrinsically, indicating the presence of thermal material interior to the SNR, e.g. unshocked ejecta, or extrinsically from the interaction of the SNR with its immediate surroundings.   Due to limited observational capabilities, intrinsic thermal absorption has only been detected in the brightest SNRs, including the Crab nebula \citep{bie97}, Cas A \citep{kas95, del14, arias+18}, and Tycho \citep{arias+19}. External thermal absorption is a tracer of the ionised interface generated as an SNR interacts with its immediate surrounding in Galactic complexes. In addition to probing this interaction, it provides constraints on the relative radial superposition of thermal and non-thermal constituents in complex regions \citep{bro03,bro05-391}. 
{\it (iii)\,}  The distribution of ionised gas in the ISM unrelated to SNRs, which can be measured using SNRs as background beacons \citep{kassim-89-list}. 
{\it (iv)\,}  The Galactic SNe rate, which is poorly known due to incompleteness in Galactic SNR catalogues.  Low frequency observations of SNRs are a proven means of addressing the incompleteness \citep{bro04, bro06, hurley19}. 

Starting in the 1990s \citep{kassim1993}, technical breakthroughs enabled a succession of dramatic improvements in low radio frequency observational capabilities \citep{vanhaarlem2013}. The scientific impacts have so far mainly been realised for extragalactic studies (e.g. \citealt{vanweeren2016,shimwell2017}), but the impact on Galactic SNRs studies is slowly starting to be felt (e.g. \citealt{bro06, supan2018, arias+18}).   The 74 MHz Very Large Array Low-frequency Sky Survey Redux (VLSSr) was the first all-sky survey to take advantage of the improved low frequency capability \citep{lan14}. As such the VLSSr established an important calibration grid for a suite of emerging new instruments. While technically limited compared to a rapidly advancing state of the art, it also remains an important resource for individual source studies. In particular, its potential for SNR studies has remained largely untapped. In this paper we present a sample of 14 bright, resolved SNRs selected from the VLSSr, which we use to address a number of the scientific issues outlined above, and to stimulate future studies as larger samples of weaker sources become accessible with modern instruments, e.g. the LBA Sky Survey (LoLSS:  \citealt{lolss+21})  with LOFAR. 

This paper is organised as follows. In Sect.~\ref{sample} we describe our selection of the 14 bright Galactic SNRs from the VLSSr images analysed in this paper. We measure their integrated 74 MHz flux densities.  When available, we also measure their low frequency flux densities from the Galactic and Extragalactic All-sky Murchison Widefield Array Survey (GLEAM, \citealt{wayth15,hurley19}). 
The method used to construct spectral index maps is explained in Sect.~\ref{local}. 
In Sect.~\ref{newly} we assimilate the new low radio frequency fluxes into a larger framework of measurements from the literature, and use it to construct improved integrated continuum spectra. Careful attention is given to anchoring these spectra on an accurate, absolute flux density scale which is valid over most of the frequency range considered for each source 
\citep{per17}. 
Together with the inclusion of new low frequency measurements the derived spectra are a marked improvement over previous studies, as explained in Sect.~\ref{literature}. In Sect.~\ref{individual} we discuss the spatially resolved morphology and spectral index behaviour for each individual SNR. In Sect.~\ref{ISM} we focus on those sources whose spectral analysis presents deviations from a canonical power law at the lowest frequencies, attributable to thermal absorption. We analyse the properties of the surrounding medium in order to understand the physical context of the absorption, i.e. whether it is intrinsic, extrinsic and proximate to the SNR, or attributable to more distant ISM from along the line of sight. We present our summary and conclusions in Sect.~\ref{summary}.

\section{General Properties of the VLSSr SNR sample}
\label{sample}
Based on their radio morphologies, our sample comprises 10 shell-type, 2 composite, and 2 plerion SNRs, all of which are previously known \citep{green19}.\footnote{\url{http://www.mrao.cam.ac.uk/surveys/snrs/}.} Three of the SNRs in the list are also part of the  mixed-morphology (MM) class showing centrally condensed thermal X-ray emission surrounded by a synchrotron radio shell. Twelve sources analysed in this work belong to the first Galactic quadrant from $l\sim$+4$^{\circ}$ to $l\sim$+44$^{\circ}$, while the remaining two objects are located in the second quadrant, in the 120$^{\circ}$ $\leq$ $l$ $\leq$131$^{\circ}$ region. All of them are located within Galactic latitudes $-$1$^{\circ}$ $\leq$ $b$ $\leq$7$^{\circ}$.

\subsection{VLSSr Data}
In Fig~\ref{74-images}, we present images for each of the 14 SNRs from the VLSSr\footnote{The latest release of the VLSSr images is available at the website \url{http://www.cv.nrao.edu/vlss/VLSSpostage.shtml}.}. The observations are centred at a frequency of $\nu=74$ MHz with an angular resolution of $\theta\sim75^{\prime\prime}$.  The images are sensitive to spatial structures up to $\sim$36$^{\prime}$ in size, larger than the full extent of any SNR in our sample.  In all cases the emission is displayed above a 4-$\sigma$ noise level measured in the corresponding VLSSr fields (the mean rms noise level of the maps is $\sim$0.16~{\Jyb}). For the majority of the SNRs in our sample, the VLSSr maps represent the most complete available image of the source, both resolving the structure and recovering the diffuse emission in the low frequency regime below 100 MHz \citep[see, for instance,][]{slee77,kassim-88}. The morphological properties of each SNR at 74~MHz are discussed below in Sect.~\ref{individual}. Throughout this paper each SNR is referred to by the name most commonly used in the literature. 
The correspondence with the name derived from the Galactic coordinates of the centre of the source is indicated in Table~\ref{74properties}.

As a qualitative measure of the consistency of our source size measurements at 74~MHz, we compared them to their counterparts at 1.4~GHz taken, depending on the sky position, from the {\sl Multi-Array Galactic Plane Imaging Survey} (MAGPIS, \citealt{helfand-06}) or the NRAO VLA Sky Survey (NVSS, \citealt{con98}).  We constrained our measurements to regions exceeding at least 3 times the respective rms noise levels in the images at 74~MHz and 1.4~GHz. In all cases the 1.4~GHz images were convolved to the VLSSr resolution (75$^{\prime\prime}$). In general the source sizes are expected to match with a few exceptions. The source could appear larger at 74~MHz than at 1.4~GHz if the higher frequency observations miss faint, extended structure \citep[see for example,][]{lan04}.  Alternately, foreground thermal absorption along the line of sight can unevenly attenuate the synchrotron emission across the SNR and cause the source to appear smaller at 74~MHz \citep{lac01}.  Finally, residual ionospheric calibration errors can distort the apparent source size at 74~MHz \citep[a further discussion of this non-physical effect is presented in][]{cohen07}. 

Despite these unknowns, our comparison indicated a remarkably good agreement in source size for 11 of our sources, with 74~MHz/1.4~GHz-size ratios $\sim1.04$. Two sources, W41 and 3C~391, have ratios of $\sim0.26$ and $\sim0.87$, respectively, indicating they are substantially smaller at 74 MHz than at 1.4~GHz.  W41 sits inside the giant molecular complex G23.3$-$0.4, and is spatially coincident with several HII regions (\citealt{mes14}, \citealt{hogge19}). As discussed in Sect.~\ref{W41}, we attribute the reduced apparent size at 74 MHz to absorption by HII regions in the complex blocking the SNR emission.   For SNR~3C~391 the result is consistent with data presented in \citet{bro05-391}, who attributed it to thermal absorption tracing the ionised interface along the SNR/molecular cloud interface.  SNR~3C~396, on the other hand, has a ratio of $\sim1.4$, indicating it is significantly more extended at 74~MHz than at 1.4~GHz.  We attribute this to the higher frequency measurements missing the very faint emission at the northwest edge.  With these three exceptions accounted for, we feel confident that our VLSSr measurements provide a robust sampling of the full SNR source sizes at 74~MHz. 

Basic radio source parameters were measured from the VLSSr images, such as the angular dimensions, the total and peak flux density, and the surface brightness of each SNR in the sample. All of them are reported in Table~\ref{74properties}. For each SNR the integrated flux density was derived by using a polygonal region to fit the outer radio boundary of the remnant 4~$\sigma$ above the intrinsic noise level measured in the corresponding VLSSr field.  When necessary, depending on the fluctuations of the background emission around the source, the flux density estimate was corrected for an average background level. This contribution was determined by scanning in both right ascension and declination through several positions surrounding the SNR.  The main errors in the listed flux densities arise from intensity-proportional flux-density uncertainties in both the absolute flux-density scale ($\sim$15\%) and primary beam corrections, as well as uncertainties in the background estimations and bias corrections. All of these contributions were combined in quadrature to compute the final error in the integrated flux density measurements. Surface brightness estimates at 74~MHz were calculated from the relation $\Sigma_{74} = 1.505 \times 10^{-19} \, S_{74}/A_{74}$~W~m$^{-2}$~Hz$^{-1}$, where $S_{74}$ is the integrated flux density (in Jy) measured in the VLSSr map and $A_{74}$ represents the  area (in square arc minutes) enclosed by the polygon region used to integrate the 74~MHz emission.  The mean percentage error in our surface brightness estimates is $\sim$30\% and is dominated by uncertainties in defining the outer boundary of the radio emission. The flux density measurements from the VLSSr maps analysed here add to the scarce list of reliable low-frequency estimates available to date.

\begin{figure*}[ht!]
  \centering
\includegraphics[width=0.85\textwidth]{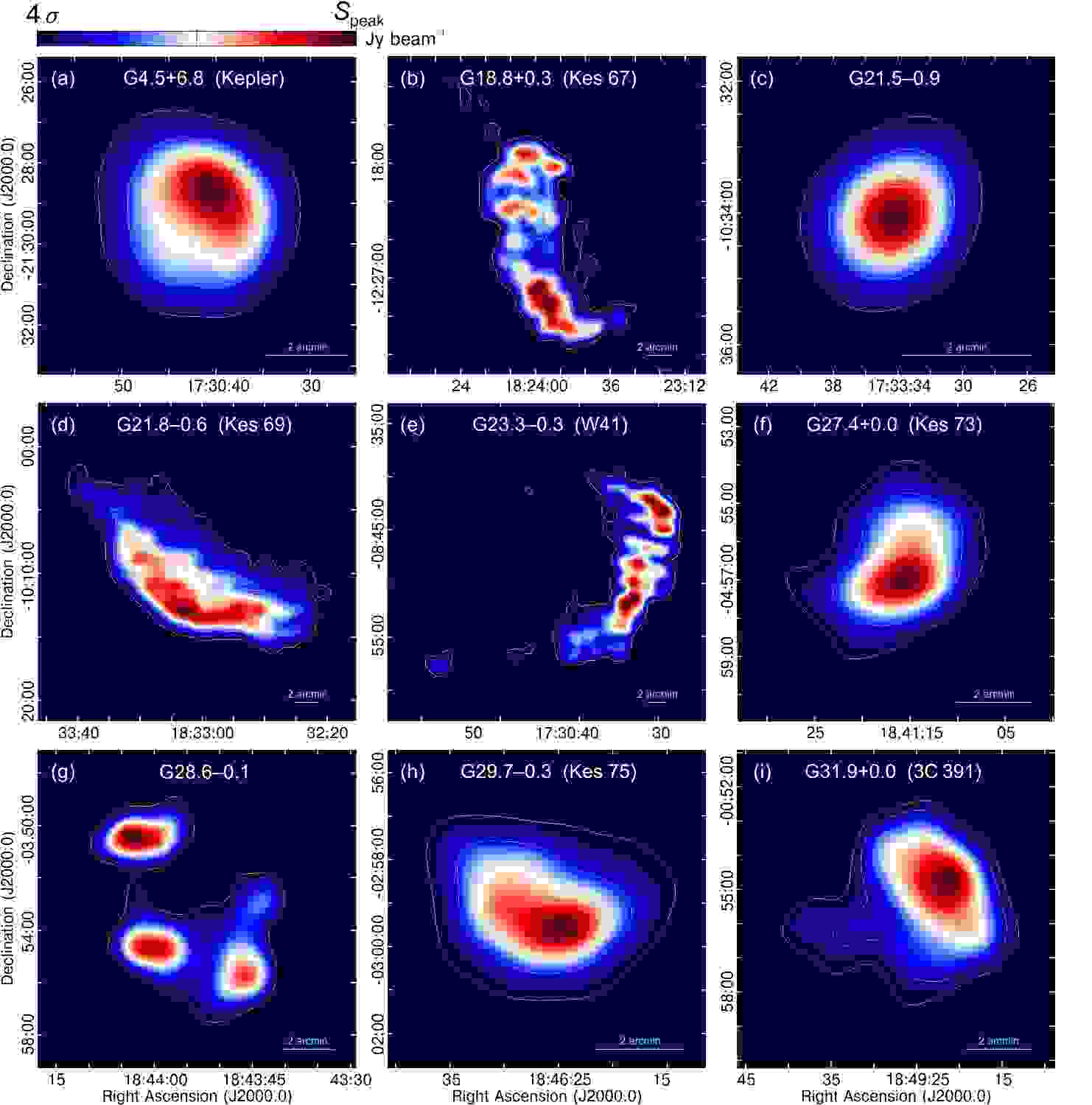}
\caption{The VLSSr 74~MHz images for the 14~SNRs in our sample, with an 75$^{\prime\prime}$ angular resolution. The colour scale, given on top, is linear scaling from 4 times the local rms noise level (4~$\sigma$) to the peak intensity value of the subimage, $S_{\mathrm{peak}}$ (see values quoted in Table~\ref{74properties}). 
The contours levels of the 74~MHz emission start at 4~$\sigma$ increasing in steps of 25, 50, and 75\% of the scale range. Exceptions are Tycho and 3C~397 for which an 8-$\sigma$ lower limit was chosen. A cyan horizontal line of 2$^{\prime}$ length is included in each panel to facilitate the comparison between the SNRs' sizes.}
 \label{74-images}
\end{figure*}

\addtocounter{figure}{-1}
\begin{figure*}[ht!]
  \centering
\includegraphics[width=0.85\textwidth]{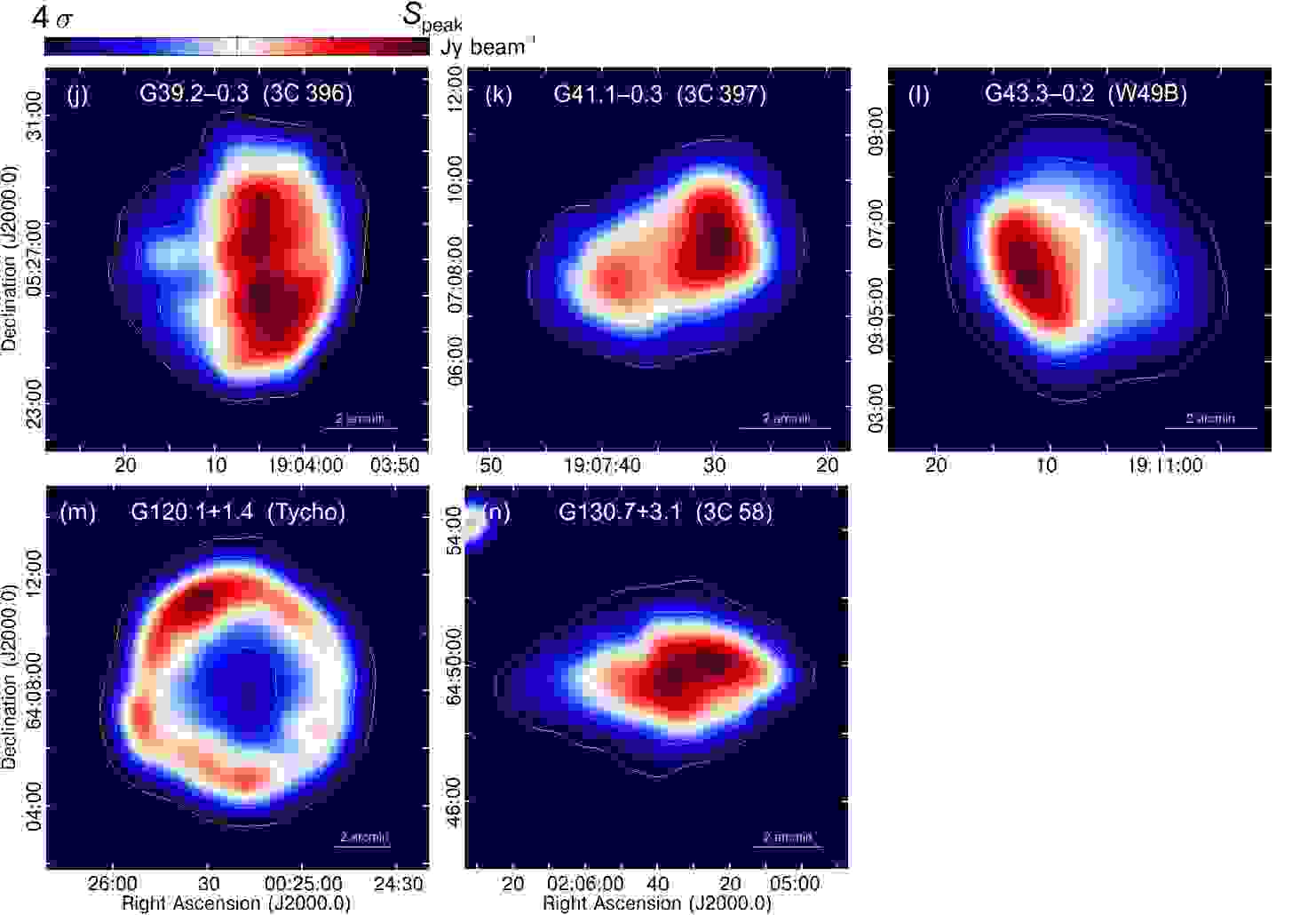}
\caption{{\itshape Continued}.}
\label{74-images}
\end{figure*}

\begin{table*}
\centering
\small
\caption{Continuum properties for all the SNRs in our sample derived from the 74~MHz VLSSr images.}
\label{74properties}
\begin{tabular}{l c c c c c c c}\hline\hline
%-----------------------------------------------------------------------------------------------------------------------------------------------
Galactic & Alternative & Morphological & Size & VLSSr Flux & $\Sigma_{74}$ & VLSSr rms & $S_{\mathrm{peak}}$\\\cline{4-4}\cline{6-6}
 name & name & Class & $\theta_\mathrm{max}$[$^\prime$] $\times$ $\theta_\mathrm{min}$[$^\prime$] & Density[Jy] & [W~m$^{-2}$~Hz$^{-1}$~sr$^{-1}$] & [{Jy~beam$^{-1}$}] & [{Jy~beam$^{-1}$}]\\\hline
%-----------------------------------------------------------------------------------------------------------------------------------------------
 \object{G4.5$+$6.8}    & \object{Kepler} &  Shell      & $5   \times 5$    &   $111\pm17$   &  $6.7\times10^{-19}$ &  0.23 & 24.4 \\ 
 \object{G18.8$+$0.3}   & \object{Kes~67} &  Shell      & $17  \times 11$   &  $76.2\pm13.8$ &  $6.1\times10^{-20}$ &  0.14 &  1.9 \\
 \object{G21.5$-$0.9}   & --              &  Plerion    & $3.2 \times 3.2$  &   $6.4\pm1.1$  &  $9.4\times10^{-20}$ &  0.15 &  4.1 \\
 \object{G21.8$-$0.6}   & \object{Kes~69} &  Shell      & $22  \times 10$   &   $169\pm31$   &  $1.2\times10^{-19}$ &  0.18 &  4.5 \\
 \object{G23.3$-$0.3}   & \object{W41}    &  Shell      & $18  \times 7$    &    $88\pm17$   &  $1.1\times10^{-19}$ &  0.14 &  2.3 \\
 \object{G27.4$+$0.0}   & \object{Kes~73} &  Shell      & $6   \times 5$    &  $13.8\pm2.5$  &  $6.9\times10^{-20}$ &  0.14 &  2.8 \\
 \object{G28.6$-$0.1}   & --              &  Shell      & $9.5 \times 7.0$  &  $26.9\pm4.7$  &  $6.4\times10^{-20}$ &  0.15 &  2.4 \\
 \object{G29.7$-$0.3}   & \object{Kes~75} &  Composite  & $4.5 \times 3.5$  &  $48.5\pm7.9$  &  $4.6\times10^{-19}$ &  0.22 & 14.0 \\
 \object{G31.9$+$0.0}   & \object{3C~391} &  MM         & $6.5 \times 5.5$  &  $31.5\pm5.4$  &  $1.2\times10^{-19}$ &  0.20 &  4.9 \\ 
 \object{G39.2$-$0.3}   & \object{3C~396} &  Composite  & $7.5 \times 7.5$  &  $44.8\pm8.8$  &  $1.2\times10^{-19}$ &  0.16 &  4.0 \\
 \object{G41.1$-$0.3}   & \object{3C~397} &  MM         & $5.5 \times4.0$   &  $68.9\pm10.6$ &  $4.7\times10^{-19}$ &  0.16 & 12.7 \\
 \object{G43.3$-$0.2}   & \object{W49B}   &  MM         & $5.5 \times5.5$   &  $64.0\pm10.1$ &  $3.2\times10^{-19}$ &  0.13 & 11.6 \\
 \object{G120.1$+$1.4}  & \object{Tycho}  &  Shell      & $9   \times 9$    & $255.0\pm38.8$ &  $4.7\times10^{-19}$ &  0.15 & 14.5 \\
 \object{G130.7$+$3.1}  & \object{3C~58}  &  Plerion    & $ 8.5\times 4.5$  &  $34.6\pm5.3$  &  $1.4\times10^{-19}$ &  0.08 &  3.7 \\ 
 \hline
 %-----------------------------------------------------------------------------------------------------------------------------------------------
\end{tabular}
\tablefoot{Columns 1 and 2 list the Galactic-coordinate names and the alias of all the SNRs included in our sample set, respectively. 
The radio morphology class of the source is noted in Column 3. If the SNR presents a mixed X-ray and radio morphology, it is indicated by the abbreviation MM. The size of each source, measured at 74~MHz from the VLSSr images, is reported in column 4. The total flux density and the surface brightness at 74~MHz are summarised in columns 5 and 6, respectively. 
Columns 7 and 8 report the rms noise level (1$\sigma$) and the peak of the 74 MHz emission calculated in the individual VLSSr maps.}
\end{table*}

\subsection{GLEAM}
In order to more fully probe the low frequency spectra of the SNRs, we also looked at recently published images from GLEAM.  This survey currently covers parts of the Galactic Plane at frequencies of 88, 118, 155, and 200 MHz, at resolutions from $4^{\prime} - 2^{\prime}$ \citep{hurley19}.  For 12 of the 14 SNRs we were able to obtain flux density measurements from GLEAM published images, and we report these measurements in Table~\ref{GLEAM}.  At the time we made the analysis, there were no images available for the remaining two remnants.  In some cases the images were of lower quality, and we chose not to use them for this study.  We measured the fluxes and errors using the same method described above for the VLSSr images.  Because the images are at a lower resolution which does not resolve many of the remnants, we present only the integrated flux measurements in this work. 

\begin{table*}
\centering
\small
\caption{Integrated flux densities measured from GLEAM Survey images of 12 SNRs in our study.}
\label{GLEAM}
\begin{tabular}{l c c c c}\hline\hline
%-----------------------------------------------------------------------------------------------------------------------------------------------
\multirow{2}{*}{Source (alias)}  & \multicolumn{4}{c}{Flux density~[Jy]} \\\cline{2-2}\cline{3-5}
  & 88~MHz & 118~MHz & 155~MHz & 200~MHz \\\hline
%-----------------------------------------------------------------------------------------------------------------------------------------------
G4.5$+$6.8~~(Kepler)   & $100  \pm 9$    & $77   \pm 7$    &  $62   \pm 9$    & ...            \\ 
G18.8$+$0.3~~(Kes~67)  & $81   \pm 17$   & $80   \pm 14$   &  $71   \pm 11$   & $62   \pm 9$   \\
G21.5$-$0.9            & $7.2  \pm 2.7$  & $6.6  \pm 2.0$  &  $6.3  \pm 1.3$  & $5.9  \pm 1.1$ \\ 
G21.8$-$0.6~~(Kes~69)  & $194  \pm 21$   & $166  \pm 20$   &  $143  \pm 16$   & $116  \pm 14$  \\
G23.3$-$0.3~~(W41)     & $154  \pm 30$   & $172  \pm 40$   &  $149  \pm 37$   & $139  \pm 28$  \\
G27.4$+$0.0~~(Kes~73)  & $17.0 \pm 3.5$  & $17.6 \pm 2.9$  &  $14.6 \pm 3.0$  & ...            \\
G28.6$-$0.1            & $29.5 \pm 5.6$  & $25.9 \pm 3.9$  &  $20.2 \pm 3.1$  & $15.5 \pm 1.9$ \\
G29.7$-$0.3~~(Kes~75)  & ...             & $42.3 \pm 5.8$  &  $32.4 \pm 3.4$  & $25.5 \pm 2.7$ \\
G31.9$+$0.0~~(3C~391)  & $43.7 \pm 8.0$  & $48.1 \pm 4.9$  &  $44.4 \pm 4.1$  & ...            \\ 
G39.2$-$0.3~~(3C~396)  &  ...            & $39.9 \pm 4.9$  &  $33.9 \pm 3.4$  & $28.5 \pm 2.8$ \\
G41.1$-$0.3~~(3C~397)  & $57   \pm 7$    &  ...            &  ...             & ...            \\
G43.3$-$0.2~~(W49B)    & $63.8 \pm 8.0$  & $69.7 \pm 5.5$  &  ...             & ...            \\\hline
%-----------------------------------------------------------------------------------------------------------------------------------------------
\end{tabular}
\tablefoot{There are no GLEAM Survey data available for SNRs G120.1$+$1.4~~(Tycho) and G130.7$+$3.1~~(3C~58). We used '...' to denote cases for which the flux density values measured from GLEAM images do not fit our selection criteria to construct radio continuum spectra, see text in Sect.~\ref{newly} for details.}
\end{table*}

\section{Local SNR variations in the radio spectral index}
\label{local}
Local changes in the radio spectral index across each source were computed for our targets by combining the VLSSr maps with the best available image of
the source from surveys at 1.4~GHz (e.g., MAGPIS, NVSS, VLA Galactic Plane Survey (VGPS, \citealt{stil06}), and the NRAO VLA Archive Survey (NVAS)\footnote{\url{httpp://www.archive.nrao.edu/nvas}}).
Since we do not have the \it uv\rm-data for the higher frequency images, we created the spectral index maps in a standard way from the direct ratio of the images at both frequencies. While doing this, we aligned, interpolated, and smoothed the 1.4~GHz maps to matching VLSSr ones. Additionally, only regions with flux densities greater than 4-$\sigma$ significance level of their respective sensitivities were used in the process. The resulting spectral index images are displayed in Fig.~\ref{alpha-maps}. Errors on these maps are of order $\sim$20-25\% for the local spectral index measurements. We are aware that a quantitative interpretation of spectral variations with position is not possible from the resulting maps, but they are very useful to reveal qualitative trends. The analysis of the radio spectral index images is presented in Sect.~\ref{individual}. We note that  for SNR 3C~397 (G41.1$-$0.3) the public radio continuum images at radio frequencies higher than 74~MHz do not recover the expected flux density accurately, and thus we decided not to create a spectral index map for this source. 
 
\begin{figure*}[ht!]
  \centering
\includegraphics[width=0.8\textwidth]{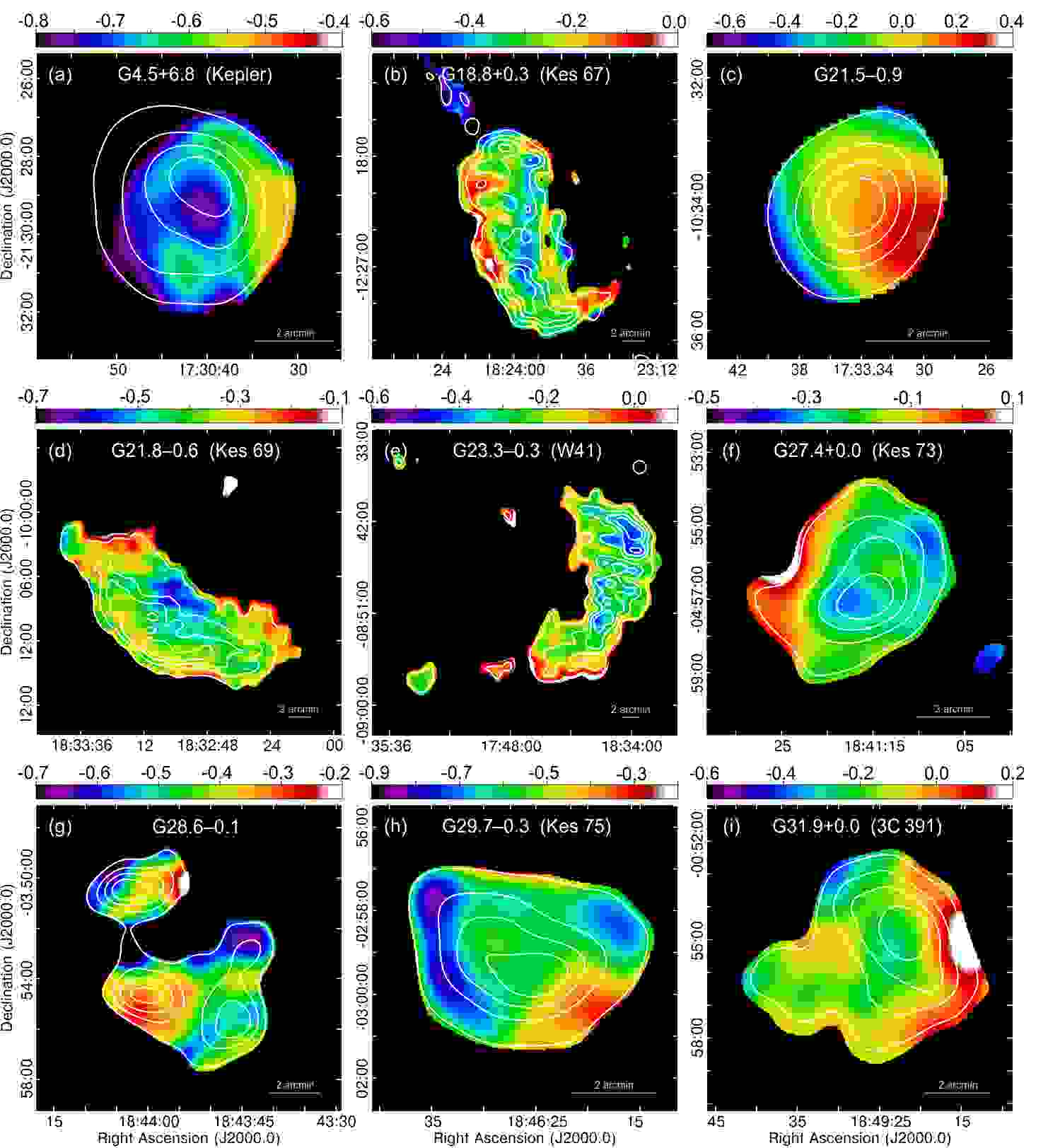}
  \caption{Spectral index maps between 74~MHz and 1.4~GHz (resolution 75$^{\prime\prime}$) for the SNRs in our sample (except the source 3C~397). 
  The maps were constructed by combining the VLSSr image with those available from radio continuum surveys (see text for details). 
  Pixels with brightness below 4$\sigma$ at 74~MHz or 1.4~GHz  were blanked. 
  The colour scales displayed at the top of the maps indicate the spectral indices measured over each SNRs. The radio continuum emission from VLSSr 74~MHz at a resolution of 75$^{\prime\prime}$ is represented by contours. For reference, we used the same contours levels as in Fig.~\ref{74-images}.  }
\label{alpha-maps}
\end{figure*}

\addtocounter{figure}{-1}
\begin{figure*}[ht!]
  \centering
 \includegraphics[width=0.8\textwidth]{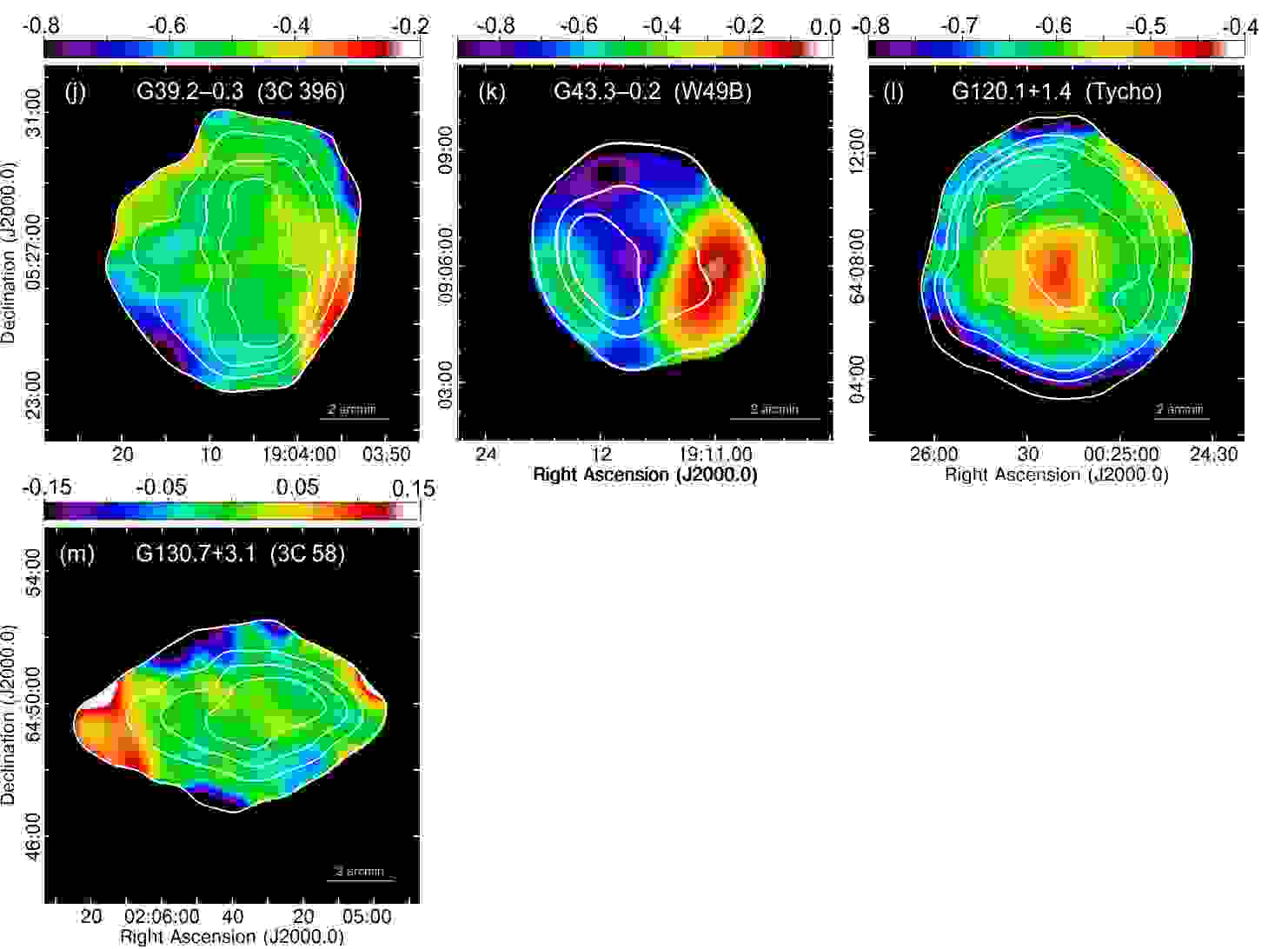}
  \caption{{\itshape Continued}}
\label{alpha-maps}
\end{figure*}

\section{Integrated SNR radio continuum spectra}
\label{fit}

\subsection{Newly derived integrated spectra}
\label{newly}

The global spectral properties of SNRs have been determined in a number of previous studies. However, the majority of them were based on the combination of data without using a common radio flux density scale, thereby making the quantitative comparison of measurements incorrect. In a few cases, the absolute scale of \citet{baars77} was used, although it was applied even at frequencies lower than 300~MHz or higher than 15~GHz for which the scale is incomplete. In addition, published spectra include observations with low angular resolution and surface brightness sensitivity, which in complex regions of the Galaxy can easily miss non-thermal components in the emission by confusion with Galactic background or thermal sources. The inclusion in the collected data of widely scattered flux density measurements at similar frequencies as well as differences in the quality of the data not correctly weighted in the analysis, also affect the reliability of previously published spectra. 

Flux densities included in our 14 integrated spectra were selected based on the  following criteria: i) we only included measurements with error estimates less than 30\%,  ii) measurements with deviations well beyond the best-fit model and inconsistent with the reported errors were excluded, iii) at frequencies above 1~GHz, we excluded interferometer measurements with insufficient short spacings to sample the extended source flux, and iv) we excluded single-dish measurements with poor resolution that may overestimate flux densities due to high confusion levels. Based on these criteria we compiled 454 flux density measurements across the frequency range of 15~MHz - 217~GHz.   We combine these with the newly measured VLSSr and GLEAM flux densities to create updated integrated radio spectra for the SNRs in our sample. The VLSSr and GLEAM points help fill in the low frequency portion of the spectra most poorly constrained by past measurements from Culgoora (80~MHz, \citealt{slee77}), Clark Lake TPT (30.9~MHz, \citealt{kassim-88}), and  the Pushchino telescopes (83~MHz, \citealt{kovalenko-94}).

When possible, the measurements were adjusted to the absolute flux density scale provided by \citet{per17}. This scale, established between 50~MHz and 50~GHz, is accurate to 3\% and up to 5\% for measurements at the extreme of the frequency range. For $\sim 20\%$ of the literature fluxes, correction was not possible because of insufficient information on primary flux calibration in the original reference, or because the frequency was outside the range for which the flux scale is defined. For these sources we included the values as reported without adjustment.  The final set of flux densities used to construct the integrated spectra are presented in Appendix~A. 

In Fig.~\ref{74-spectra} we present the radio continuum spectra for all SNRs in our study.  VLSSr and GLEAM flux densities are indicated by filled red and yellow symbols, respectively, and all points are weighted by their estimated uncertainties. In each spectrum the 1- and 2-$\sigma$  error in the best-fit values is represented by gray-shaded regions. We found that 5 of the SNRs could be fit by power law functions defined by the relation $S_{\nu} \propto \nu^{\alpha}$, in which $S_{\nu}$ denotes the flux density at the frequency $\nu$. This includes the two pulsar wind nebulae (PWNe) in our sample, whose spectra were fit by broken power laws. The remaining cases show evidence of absorption below 100~MHz \citep{kas89s}, and we fit the spectra with a power law and an exponential turnover, according to Eq.~\ref{turnover},

\begin{equation}
  S_{\nu}=S_{\nu_{0}}\, \left(\frac{\nu}{\nu_{0}}\right)^{\alpha} \, 
  \mathrm{exp}\left[-\tau_{\nu_{0}}\,\left(\frac{\nu}{\nu_{0}}\right)^{-2.1}\right].
\label{turnover}
\end{equation}

\noindent
where $\nu_{0}$ is the reference frequency, set to 74~MHz, at which the integrated flux density $S_{\nu_{0}}$ and optical depth $\tau_{\nu_{0}}$ are measured.  

This is a simplistic fitting model, and we note that there are theoretical grounds to expect intrinsically curved SNR spectra, both spatially resolved and integrated. Concave-up curvature has been linked to non-linear acceleration processes in young SNRs (e.g. Tycho, Kepler; \citealt{reynolds92}) (see also Sect.~\ref{individual}). Conversely, concave-down spectra have been linked to bends in the energy distribution of the radiating electrons \citep{anderson93}. Also, synchrotron losses, thermal bremsstrahlung, and spinning dust in evolved SNRs (e.g. 3C~391 and 3C~396) interacting with high density environments have been proposed to flatten spectra at higher frequencies ($\sim$10-100~GHz) \citep{urosevic2014}. Since we find no compelling evidence for these signatures in our spectra, we proceeded with the power law plus thermal absorption model. There is still tremendous room for significantly improved measurements, at both high and low frequencies, that may well eventually justify more complex modelling for many of these sources.  

Table~\ref{table2} provides a summary of the best-fit spectral indices and free-free optical depths (when appropriate) from the weighted fit to the integrated continuum spectra shown in Fig.~\ref{74-spectra}. We also include any values for these parameters previously published in the literature.  The new results are used in Sect.~\ref{ISM} to constrain the physical properties (electron measure, $\mathrm{EM}$, and electron density, $n_{\mathrm{e}}$) of the foreground ionised gas responsible for the measured spectral turnovers. 

In Fig.~\ref{histogram}a we have plotted the distribution of the integrated radio spectral indices from our power-law fits. For comparison, the age and morphology of each source is also indicated in Fig.~\ref{histogram}b. For our purpose, we adopted the usual classification of ``young'' to refer to a SNR in either the free-expansion or the early Sedov phase of its evolution ($\lesssim$3000~yr). SNRs in subsequent evolutionary stages are considered evolved, an admitted simplification as multiple evolutionary phases may occur simultaneously in different parts of a SNR. We found that 8 sources have relatively steep spectra ($|\alpha|>0.5$), 5 of which are young. In addition, there are 4 SNRs with flatter radio spectra ($|\alpha|<0.5$). The remaining two sources are the pulsar wind nebulae G21.5$-$0.9 and 3C~58, for which the continual injection of energetic electrons produces even flatter integrated spectra. These results are further discussed in Sect.~\ref{individual}.  

The spectral indices that we have measured in this work for young and evolved SNRs disagree with test particle predictions from diffusive shock acceleration theory  \citep{reynolds12-alpha,urosevic2014}. In the linear regime, flatter spectral indices (the flattest possible value is $\alpha=-0.5$) are predicted for parallel shocks in the most energetic young SNRs, while steeper values are expected for older objects with much lower shock velocities. Explanations for the gradual flattening of the radio spectra with aging of SNRs, or alternatively, the steeper spectra of young objects, include oblique-shocks \citep{bell11}, Alfv\'enic drift effect in the downstream and/or upstream regions of the forward shock (e.g. \citealt{jiang13,slane2014}), 
shock acceleration with particle feedback \citep{pavlovic17}, and turbulent magnetic field amplification \citep{bell19}.  Contamination by intrinsic thermal bremsstrahlung radiation and high compression ratios for radiative shocks are also expected to contribute to the flat spectra observed in intermediate-age and evolved SNRs \citep{onic13}.  Individual discussions on the spectral properties of the sources in our sample are presented in Sect.~\ref{individual}.

\begin{figure*}[h!]
\centering
\includegraphics[width=0.75\textwidth]{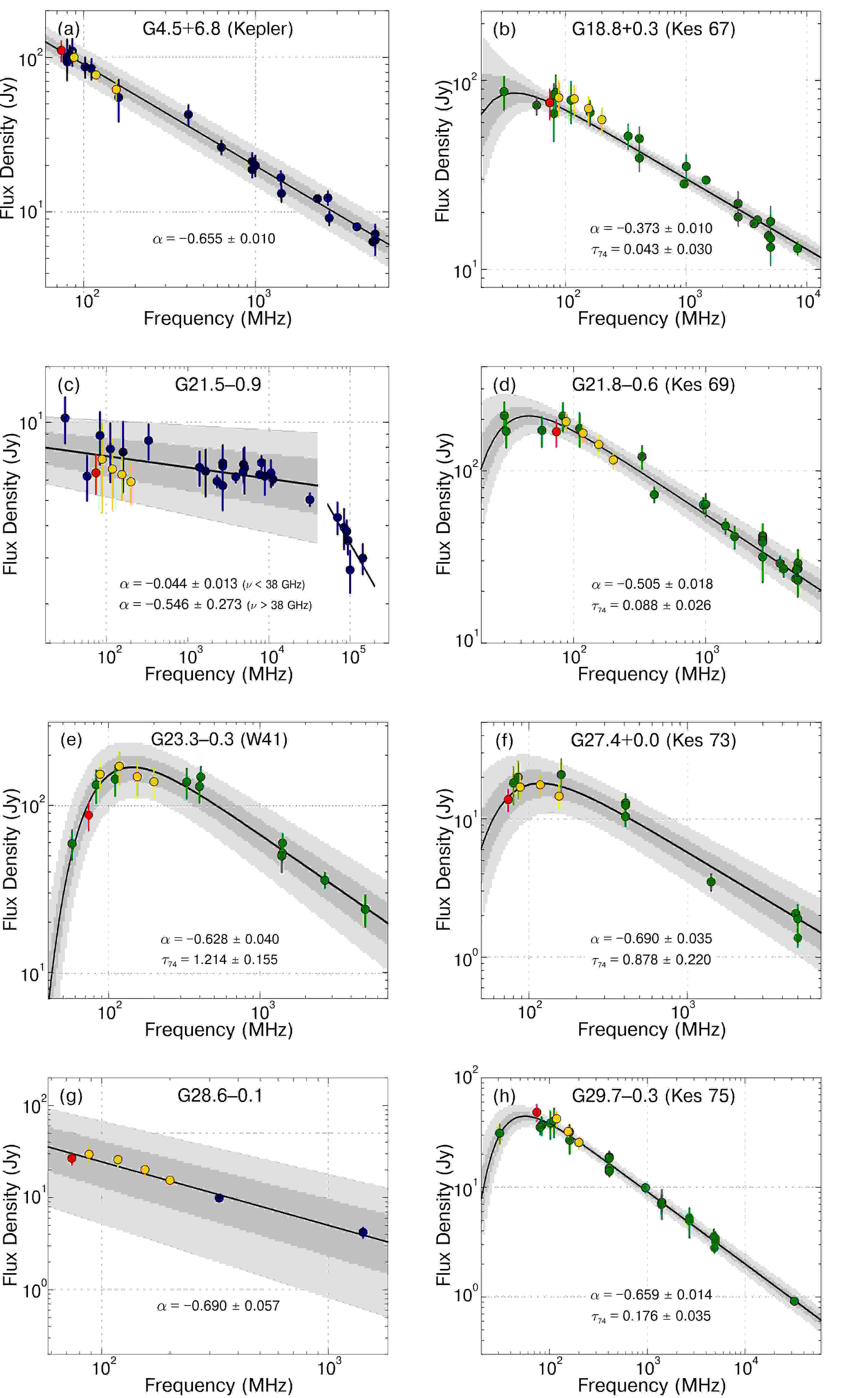}
\caption{Revised integrated radio continuum spectra for the 14 SNRs in the VLSSr sample. In each spectrum the red filled circle indicates the new 74~MHz VLSSr flux density measurements, and the yellow ones the new GLEAM measurements. The remaining values are taken from the literature and plotted in blue or green depending on whether a single power law or a power law with a low-frequency turnover model was used to fit the data (see text for details). The solid line represents the best-fitting curve to the weighted data. Measurements were adjusted to the absolute flux density scale of  \citet{per17}. Gray-shaded bands represent the 1- and 2-$\sigma$ statistical uncertainty in the best-fit values of the spectral index $\alpha$ and the free-free optical depth $\tau_{74}$ (indicated in the lower portion of each panel, see also Table~\ref{table2}).} 
\label{74-spectra}
\end{figure*}

\addtocounter{figure}{-1}
\begin{figure*}[h!]
  \centering
\includegraphics[width=0.75\textwidth]{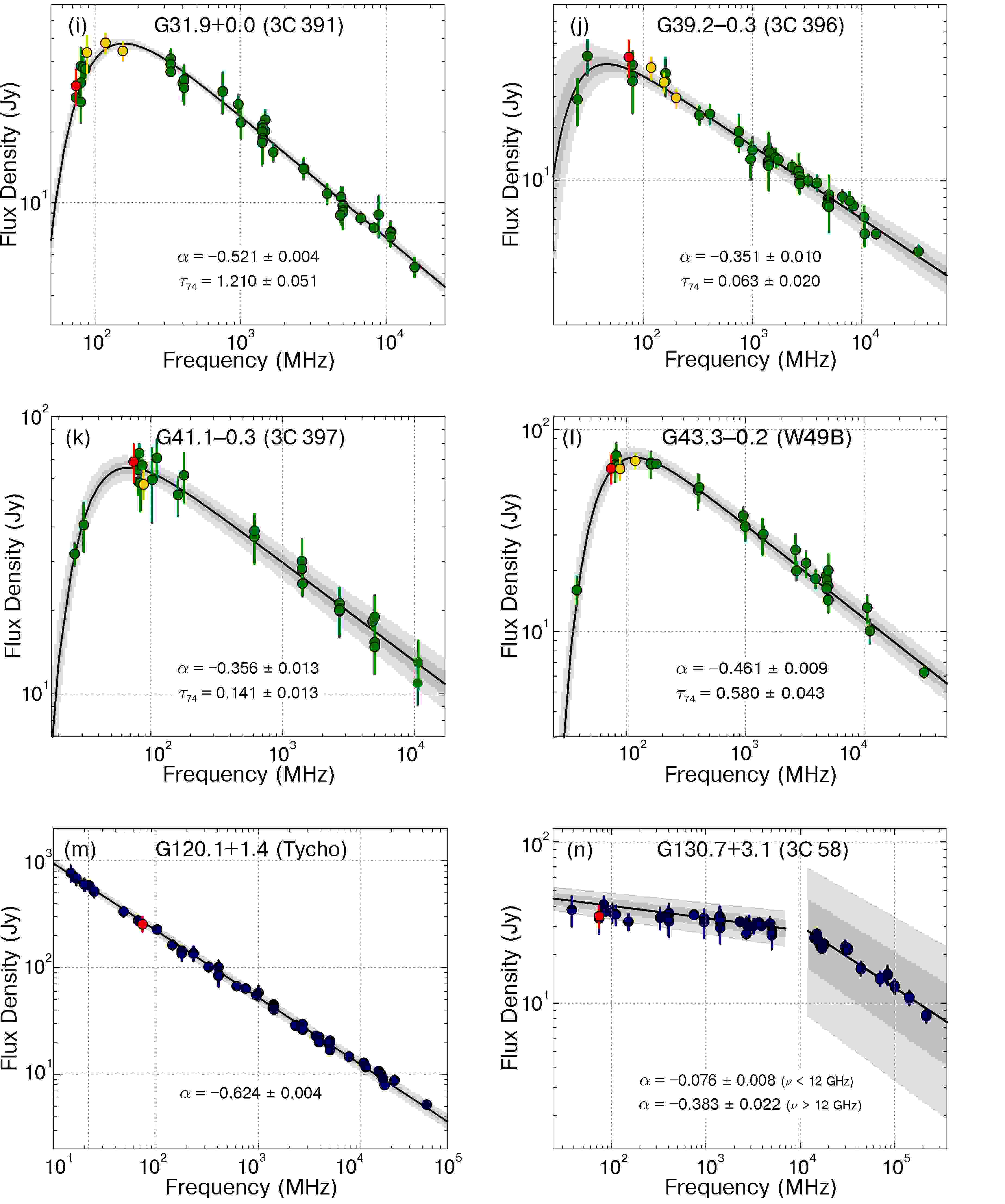}
\caption{{\itshape Continued}.}
  \label{74-spectra}
\end{figure*}

\begin{figure}[ht!]
  \centering
\includegraphics[width=0.45\textwidth]{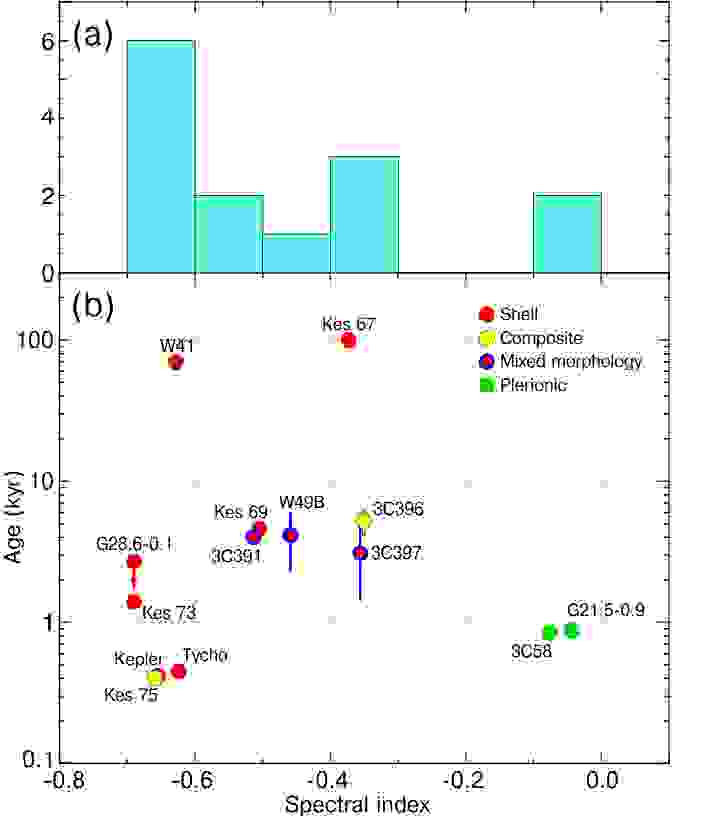}
\caption{\it Panel a: \rm Distribution of the radio continuum spectral indices as inferred from the weighted fitting to the entire spectrum of each SNR in our study (Fig.~\ref{74-spectra}). \it Panel b: \rm Distribution of the computed spectral index values according to the age and morphology of the source. The spectral indices for the younger ($\lesssim$3000~yr) SNRs are steeper than $-$0.5, the canonical value in test-particle diffusive shock acceleration theory.}
 \label{histogram}
\end{figure}

%%%%%%%%%%%%%%%%%%%%%%%%%%%%%%%%%%%%%%%%%%%%%%%%%%%%%%%%%%%%%%%%%%%%%%%%%%%%%%%%%%%%%%%%%%%%%%%%%%%%%%%%%%%%%%%%%%%%%%%%%%%%%%%%%%%%
\begin{table*}
\centering
\small
\caption{Radio continuum spectral index, free-free optical depth at 74~MHz, and ISM properties -- emission measures ($\mathrm{EM}$) and electronic densities ($n_\mathrm{e}$) --  computed from our fit to the spectra presented in Fig.~\ref{74-spectra}.}
\label{table2}
\begin{tabular}{lcccccccc}
%-----------------------------------------------------------------------------------------------------------------------------------
 \multicolumn{8}{c}{{\bfseries Absorption  from extended envelopes of normal HII regions (EHEs)}}\\\hline\hline 
%-----------------------------------------------------------------------------------------------------------------------------------
  \multirow{3}{*}{Source (alias)}  & \multicolumn{4}{c}{New results}  && \multicolumn{3}{c}{Literature} \\\cline{2-5}\cline{7-9}
  & \multirow{2}{*}{$\alpha$} & \multirow{2}{*}{$\tau_{74}$} & $\mathrm{EM}$\tablefootmark{~a} & $n_\mathrm{e}$\tablefootmark{~a} & & \multirow{2}{*}{$\alpha$\tablefootmark{~b}} & \multirow{2}{*}{$\tau_{74}$\tablefootmark{~c}} & \multirow{2}{*}{Refs.}    \\
  &  &  & [cm$^{-6}$~pc] & [cm$^{-3}$] &  &  & \\\hline\hline
%-----------------------------------------------------------------------------------------------------------------------------------
 G18.8 $+$ 0.3~~(Kes~67) &  $-0.373 \pm 0.010$  &  $0.043 \pm 0.030$  &  $150 \pm 130$  &  $1.2 \pm 0.6$  && $-0.46$ - $-0.42$  & $0.06 \pm 0.03$  & 1, 2; 1    \\
 G21.8 $-$ 0.6~~(Kes~69) &  $-0.505 \pm 0.018$  &  $0.088 \pm 0.026$  &  $440 \pm 140$  &  $2.1 \pm 0.3$  && $-0.56$ - $-0.50$  & $0.11 \pm 0.04$  & 1, 2; 1    \\
 G29.7 $-$ 0.3~~(Kes~75) &  $-0.659 \pm 0.014$  &  $0.176 \pm 0.035$  &  $870 \pm 190$  &  $2.9 \pm 0.3$  && $-0.73$ - $-0.59$  & $0.11 \pm 0.04$  & 1, 3; 1    \\
 G41.1 $-$ 0.3~~(3C~397) &  $-0.356 \pm 0.013$  &  $0.141 \pm 0.013$  &  $720 \pm  70$  &  $2.7 \pm 0.2$  && $-0.59$ - $-0.46$  & $0.13 \pm 0.01$  & 1, 2, 4; 1 \\
  &  &  &  &  &  &  &  &  \\
%-----------------------------------------------------------------------------------------------------------------------------------
 \multicolumn{8}{c}{\bfseries Absorption: Special cases\tablefootmark{~d}} \\\hline\hline
%-----------------------------------------------------------------------------------------------------------------------------------
  \multirow{3}{*}{Source (alias)}  & \multicolumn{4}{c}{New results}  && \multicolumn{3}{c}{Literature} \\\cline{2-5}\cline{7-9}
  & \multirow{2}{*}{$\alpha$} & \multirow{2}{*}{$\tau_{74}$} & $\mathrm{EM}$ & $n_\mathrm{e}$ & & \multirow{2}{*}{$\alpha$\tablefootmark{~b}} & \multirow{2}{*}{$\tau_{74}$\tablefootmark{~c}} & \multirow{2}{*}{Refs.} \\
  &  &  & [cm$^{-6}$~pc] & [cm$^{-3}$] &  &  & \\\hline\hline
%-----------------------------------------------------------------------------------------------------------------------------------
 G23.3 $-$ 0.3~~(W41)     & $-0.628 \pm 0.040$ & $1.214 \pm 0.155$ & 10     $\times$ 10$^{3}$     & 60 && $-0.63$ - $-0.48$  & $1.04 \pm 0.11$   & 1, 5; 1 \\
 G27.4 $+$ 0.0~~(Kes~73)  & $-0.690 \pm 0.035$ & $0.878 \pm 0.220$ & 8 - 15 $\times$ 10$^{3}$     & 70 - 100 && $-0.71 \pm 0.11$   & $0.72 \pm 0.32$   & 1; 1 \\
 G31.9 $+$ 0.0~~(3C~391)  & $-0.521 \pm 0.004$ & $1.210 \pm 0.051$ & 600 - 2500\tablefootmark{~e} & 10 - 40\tablefootmark{~e} & & $-0.54$ - $-0.49$    & $1.1$ & 1, 2, 6; 6 \\
 G39.2 $-$ 0.3~~(3C~396)  & $-0.351 \pm 0.010$ & $0.063 \pm 0.020$ & ---                          & --- && $-0.53$ - $-0.34$ & $0.12 \pm 0.04$   & 1, 2, 4, 7; 1  \\
 G43.3 $-$ 0.2~~(W49B)    & $-0.461 \pm 0.009$ & $0.580 \pm 0.043$ & 6 - 19 $\times$ 10$^{3}$     & $500$\tablefootmark{~f}  && $\simeq${$-0.47$}   & $0.14 \pm 0.05$   & 1, 2; 8 \\
  &  &  &  &  &  &  \\
%-----------------------------------------------------------------------------------------------------------------------------------
\end{tabular}
%-----------------------------------------------------------------------------------------------------------------------------------
\begin{tabular}{lcccc}
\multicolumn{5}{c}{\bfseries No absorption} \\\hline\hline
%-----------------------------------------------------------------------------------------------------------------------------------
  \multirow{2}{*}{Source (alias)}  & New results && \multicolumn{2}{c}{Literature}          \\\cline{2-2}\cline{4-5}
                                   & $\alpha$    && $\alpha$\tablefootmark{~b}                       & Refs.  \\\hline\hline
%-----------------------------------------------------------------------------------------------------------------------------------
 G4.5 $+$ 6.8 ~~(Kepler)           &   $-0.655\pm0.010$      && $-0.65$ - $-0.62$ & 1, 9                \\
 G21.5 $-$ 0.9 
 &  $\left\{\begin{array}{c}\hspace{-1.5mm}-0.044\pm0.013  
 \hspace{0.3cm} \nu \lesssim 38~\mathrm{GHz}          \\
        \hspace{-1.8mm}-0.546\pm0.273 
                       \hspace{0.3cm} \nu > 38~\mathrm{GHz}                     \end{array}
                                              \right.$
                                           && $~~\begin{array}{c}\hspace{-1.5mm} -0.09~\mbox{-}~{+}0.08 \\
                                              \hspace{-1.8mm}-0.57~\hbox{-}~{-}0.37\end{array}$ &  
                                              $~~\begin{array}{c}\hspace{-1.5mm}                         1, 2, 10, 11, 12 \\
                                              \hspace{-1.8mm}                                            4, 12             \end{array}$ \\
 G28.6 $-$ 0.1                             &  $-0.690  \pm 0.057$ && $-0.6$ - $-0.5$ &                   13               \\
 G120.1 $+$ 1.4~~(Tycho)                   &  $-0.624  \pm 0.004$ && $-0.65$ - $-0.58$                   & 1, 2, 9, 14, 15              \\
 G130.7 $+$ 3.1~~(3C~58) 
 & $\left\{\begin{array}{c}\hspace{-1.5mm}-0.076\pm0.008  \hspace{0.3cm} \nu \lesssim 12~\mathrm{GHz}   \\
\hspace{-1.5mm}-0.383\pm 0.022
\hspace{0.3cm} \nu > 12~\mathrm{GHz}\end{array}
                                              \right.$
                                           && $~~\begin{array}{c}\hspace{-1.5mm} -0.10~\hbox{-}~{-}0.04                           \\
                                              \hspace{-1.8mm} -0.58~\hbox{-}~{-}0.45 \end{array}$        & 
                                              $~~\begin{array}{c}\hspace{-1.5mm}                         1, 2, 11, 14, 16, 17     \\
                                              \hspace{-1.8mm} 17, 18 \end{array}$ \\\hline
\end{tabular}
%-----------------------------------------------------------------------------------------------------------------------------------
\tablefoot{
\tablefoottext{a}{For EHEs, $\mathrm{EM}$ and $n_{\mathrm{e}}$ calculations were done using a typical electron temperature of 5000 K and an average path length of 100 pc \citep{kas89s}.} \\
\tablefoottext{b}{Numbers without errors indicate that uncertainties are not given in the references}.\\
\tablefoottext{c}{Free-free continuum optical depth reported in the literature at a frequency $\nu$ were extrapolated to 74~MHz according to 
$\tau_{74}=\tau_{\nu}\,[74~\mathrm{MHz}/\nu~(\mathrm{MHz})]^{-2.1}$.}\\
\tablefoottext{d}{The properties of the absorbing medium  are analysed  in Sect.~\ref{specials}.}\\
\tablefoottext{e}{The listed $\mathrm{EM}$ and $n_{e}$ values were extracted from the analysis in \citet{bro05-391}.}\\
\tablefoottext{f}{The estimate $n_{e}$=500~cm$^{-3}$ comes from \citet{zhu+14} for the  postshock near-IR line emitting gas.}\\
{\bfseries References.} 
 (1)  \citet{kovalenko-94}, 
 (2)  \citet{sun11}, 
 (3)  \citet{becker-helfand-84}, 
 (4)  \citet{anderson93}, 
 (5)  \citet{trushkin-98}, 
 (6)  \citet{bro05-391},
 (7)  \citet{cru16},
 (8)  \citet{kassim-89-list},
 (9) \citet{reynolds92},
 (10) \citet{bietenholz-bartel-08},
 (11) \citet{salter-89b},
 (12) \citet{iva19-g21},
 (13) \citet{helfand89},
 (14) \citet{kothes-06},
 (15) \citet{arias+19},
 (16) \citet{bie01},
 (17) \citet{iva19-3c58},
 (18) \citet{green-92}}
\end{table*}
%%%%%%%%%%%%%%%%%%%%%%%%%%%%%%%%%%%%%%%%%%%%%%%%%%%%%%%%%%%%%%%%%%%%%%%%%%%%%%%%%%%%%%%%%%%%%%%%%%%%%%%%%%%%%%%%%%%%%%%%%%%%%%%%%%%%

\subsection{Comparison to the literature}
\label{literature}
Two large samples of integrated spectra for Galactic SNRs may be found in the literature, both with flux density measurements on the  absolute scale of \citet{baars77}. \citet{kovalenko-94-spec} (hereinafter referred to as K94) compiled spectra for 102 SNRs and \citet{sun11} (S11) compiled spectra for 50 sources. Before presenting the individual analysis for the sources in our list, we first examine general differences and agreements between the spectral properties reported in these two works and ours.  The K94 sample includes spectra for 13 of the 14 SNRs in our sample, and the S11 sample includes 10 SNRs from our list. Although both works include the two PWNe from our VLSSr sample, K94 did not consider any spectral breaks in the spectrum, and S11 placed the break points for fitting the high and low frequency spectral components at different frequencies than our analysis. This makes comparisons to either work difficult, and we exclude them in the subsequent discussion. Among the remaining 11 sources which appear in both the K94 and VLSSr samples, we have found a significant discrepancy in the spectral indices for 3C~396 and 3C~397.  Our analysis yields flatter values for both by $\sim0.1$, which is 10 times larger than  the uncertainty in our spectral fits for these sources.  Spectral indices for the rest of the sources agree within the errors of the two measurements.  This is partly due to the relatively large errors (typically 7-25\%) reported on the spectral indices estimated by K94.

Among the eight sources (excluding the PWNe G21.5$-$0.9 and 3C~58) which appear in both the S11 and VLSSr samples, three of the S11 spectral indices match our new values within the reported errors (SNRs Kes~75, 3C~396, and W49B), while the rest of their determinations (SNRs Kes~67, Kes~69, 3C~391, 3C~397, and Tycho) show differences with ours. 
We notice that for 3C~391 the disagreement between our result and that of these authors arises from the fact that they modelled the spectrum of the source with a break at $\nu \sim 1$~GHz. We also highlight that the spectra in S11 are limited to measurements at frequencies $\nu > 180$~MHz. This avoids the identification of low frequency turnovers, which can be used to probe the properties of the ionised gas. Concerning the errors in S11 spectral indices, they are comparable to those we have found in this work and range from $\sim$2 to 5\%.

Details of the literature values for specific sources are included in the individual source discussions in the next section.
\rm

\section{Image and spectral analysis of individual objects}
\label{individual}
In this section, we focus on the surface brightness distribution revealed in the spatially resolved VLSSr images of our sample, interpreted in the context of  their integrated and local continuum spectra. 

\vspace{0.3cm}
\noindent
\bf Kepler's SNR (G4.5+6.8): \rm 
Even though numerous multi-wavelength studies from the optical to the X-ray bands have been published on this remnant, its properties at low radio frequencies remain poorly explored to date \citep[][and references therein]{sankrit16}. In the VLSSr image (Fig.~\ref{74-images}a) Kepler's SNR consists of a roughly spherical shell structure of about 5$^{\prime}$ in size. The highest emissivity comes from the northwest and accounts for about 40\% of the total flux density at 74~MHz ($S_{74} = 111 \pm 17$~Jy, Table\ref{74properties}). In contrast to higher frequency observations, a ridge of emission connecting the southeast region with the central part of the SNR is not evident at 74~MHz \citep{mat84,Delaney02}. 
Although the eastern and western `ears'' observed at $\sim$1.4 and $\sim$4.8~GHz \citep{Delaney02} appear to be missing, this is due to the lower resolution of the 74 MHz image, which is not sufficient to distinguish them.  Emission in these regions is detected at 4~$\sigma$ ($\sim$~0.9~Jy~beam$^{-1}$) significance.  

Fig.~\ref{alpha-maps}a  shows the spectral index map for  SNR~Kepler that we have created by combining the 74~MHz-VLSSr and the 1.4~GHz NVSS images. The spectral index values range from $\sim-0.53$ to $\sim-0.75$, which is  consistent with what was reported by \citet{Delaney02} between 1.5 and 5~GHz. In our low-frequency spectral map the flattest indices are measured in the western side of Kepler~SNR.

A single power law slope $\alpha=-0.655\pm0.010$ adequately fits the compiled flux densities measured between 74~MHz and 5~GHz (Fig.~\ref{74-spectra}a). This result, consistent within uncertainties with that from \citet{reynolds92} and with what \citet{Delaney02} found above 1~GHz, contradicts the natural expectation from test particle calculations for flat-spectrum emission produced by fast shocks in a young object where either non-linear effects (e.g. \citealt{reynolds92}, \citealt{ferrand-efficient+2014}) or quasi-perpendicular magnetic field configurations become important 
(e.g., \citealt{ferrand-perpendicular+14}). 

\citet{reynolds92} fit the integrated radio spectrum of Kepler SNR using a non-linear shock model with a small positive curvature that gradually flattens from $\sim$30 MHz to 10 GHz ($\alpha_{<1\mathrm{GHz}} = -0.684 \pm 0.024$ and $\alpha_{>1\mathrm{GHz}} = -0.586 \pm 0.063$). The data used in the \citet{reynolds92} spectrum include a point at 30~MHz for which a substantially large flux excess is observed, when compared with the value obtained by extrapolating the higher radio frequency estimates. This measurement, originally reported by \citet{jones74}, is the continuum peak flux density per synthesised beam for Kepler taken from an aperture synthesis survey of the Galactic plane carried out with the Fleurs observatory. The data have relatively low sensitivity, and an angular resolution ($\sim$~$0.^{\circ}8$), which is much larger than the $5^\prime$ source diameter. We thus feel that the measurement has a high probability of being overestimated, and have chosen to exclude it from the literature data used to construct our new spectrum, shown in Fig.~\ref{74-spectra}a. We find $\alpha_{<1\mathrm{GHz}} = -0.627\pm0.018$ and $\alpha_{>1\mathrm{GHz}} = -0.753\pm0.046$ from the fits to our new spectrum for frequencies below and above 1~GHz, respectively.  Although the higher frequency spectral index is slightly steeper, we do not feel there is convincing evidence for a spectral curvature based on the available data.  More high quality data at the lowest frequencies are needed to better define any potential curvature.

\vspace{0.5cm}
\noindent
\bf SNR Kes~67 (G18.8+0.3): \rm 
At 74~MHz this source has an elongated shape with a major axis length of $\sim$17$^{\prime}$ and a mean width of 14$^{\prime}$ (see Fig.~\ref{74-images}b). The brightest emission at 74~MHz occurs on its eastern periphery. Overall the low-frequency surface brightness distribution resembles that observed at 1.4~GHz using multiple configurations of the VLA \citep{dubner-96}. We note a plume of faint non-thermal emission extends beyond the northeast part of the shell, and is observed at a 4-$\sigma$ noise level in the VLSSr map only. The reality of this feature is questionable without further observations, and we did not include it when estimating the integrated flux density at 74~MHz listed in Table~\ref{74properties}.\footnote{The flux density of the plume in Kes~67 measured on the VLSSr image is $\sim$5~Jy, less than 7\% of the integrated SNR flux density reported in Table~\ref{74properties}.}

We have analysed variations in the local spectral index of the radio continuum emission from Kes~67 using the 74-MHz VLSSr and the 1.4~GHz MAGPIS images (see Fig.~\ref{alpha-maps}b). The measured values range from $\sim-0.15$ to $\sim-0.4$. The flattening dominating the northeastern border is consistent with a blast wave running into molecular material \citep{dubner-04}, while the values towards the southeastern portion seem to align with HII regions  \citep{paron12}. 

We fitted the integrated flux density values between 30.9~MHz and 8.4~GHz using a power law with an exponential turnover, shown in Fig.~\ref{74-spectra}b, obtaining a spectral index $\alpha=-0.373\pm0.010$. This value is consistent, within errors, with that measured between the same range of frequencies by K94 ($\alpha=-0.42\pm0.11$), but has a significantly lower error.  On the other hand, S11 used a pure power law with a slope of $\alpha=-0.46\pm0.02$ to fit the spectrum of this source from 330~MHz to 8.4~GHz, which is steeper than both our fit and that of K94.   The discrepancy points to the importance of low frequency measurements in accurate spectral fits. 

\vspace{0.5cm}
\noindent
\bf SNR G21.5$-$0.9: \rm 
Emission from this well-known Crab-like SNR has been detected in radio, infrared, and X-ray bands \citep[see, for instance][]{bietenholz-11,zajczyk12,ninka14, hitomi2018}. 
First discovered in the 1970's \citep{wilson76}, the VLSSr and the GLEAM images are the highest quality that have been published for this source at frequencies below 330~MHz. 
As revealed in Fig.~\ref{74-images}c, at low-radio frequencies the source has an elliptical structure with the axis of symmetry running approximately 30$^{\circ}$ clockwise from the north-south direction, while radio imaging at 5~GHz \citep{bietenholz-11} and in X-rays \citep{mat10} show that this elongation runs the same amount in the opposite direction. 
At higher radio frequencies and X rays, G21.5$-$0.9 has an irregular structure with granular and patchy features. These fine spatial structures are not resolved at the VLSSr angular resolution. 

The spectral index image in Fig.~\ref{alpha-maps}c was constructed from maps of the radio emission at 74~MHz and 1.4~GHz from VLSSr and NVAS, respectively. It shows a relatively uniform distribution over G21.5$-$0.9. The spectral inversion in the southwest region could represent a signature of thermal absorption from [FeII] 1.64~$\mu$m line-emitting material detected behind the shock \citep{zajczyk12}. 

Figure~\ref{74-spectra}c shows the most complete version of the G21.5$-$0.9 integrated synchrotron radio spectrum presented to date, along with a broken power-law fit. A power law with slope $\alpha=-0.044\pm0.013$  matches the flat spectrum of the photons radiating from $\sim$57~MHz to 32~GHz.  The spectrum becomes considerably steeper $\alpha=-0.546\pm0.273$ at higher frequencies. Our analysis indicates that the break occurs at a frequency near 38~GHz, though the spectrum is poorly sampled between $\sim$11 and 70~GHz. The only detection reported in this range is at 32~GHz. The gap between the lower and higher portion in the radio continuum spectrum of G21.5$-$0.9 complicates a precise determination of the spectral break. Previously, using only a data point at 1~GHz together with microwave measurements, \citet{planck2016} reported a break frequency at 40~GHz, which they associated with synchrotron cooling in a source with continuous energy injection. 
More recently \citet{xu+19}, using a limited number of radio fluxes together with X-ray estimates, claimed evidence for spectral steepening at 50~GHz. They attributed this spectral shape to two competing mechanisms, adiabatic stochastic acceleration (ASA) and synchrotron cooling, at low (in the radio band) and high (X-ray band) energies, respectively. New radio continuum observations filling the gap at intermediate radio frequencies are required to properly determine the spectral form. A precise determination of the spectral break in conjunction with an age estimate could constrain the magnetic field strength, independent of equipartition assumptions.

\vspace{0.5cm}
\noindent
\bf SNR Kes~69 (G21.8$-$0.6): \rm 
This remnant is an incomplete shell at 74~MHz with the surface brightness fading from the eastern to the western side of the remnant (see Fig.~\ref{74-images}d). 
The HII region G021.884$-$00.318 towards the northwest of the SNR shell \citep{and14} appears in absorption in the VLSSr map, which is expected for this kind of object as they become optically thick at low radio frequencies against the Galactic non-thermal background \citep{nord2006}. 
To our knowledge, no electron temperatures have been reported for the thermal source. Working constraints can be obtained through the relation presented in  \citet{quireza06}, derived from an electron temperature gradient in the Galactic disk,  $T_\mathrm{e} = (5780\pm350) + (287\pm46)\,R_\mathrm{Gal}$, where $R_\mathrm{Gal}$ is the Galactocentric distance to the HII region. 
For G021.884$-$00.318 placed at $\sim$10.7~kpc \citep{and14}, we have  $R_\mathrm{Gal} \simeq 4.2$~kpc, and thus this implies a characteristic $T_\mathrm{e} \simeq 7000$~K. For an optically thick HII region, the cosmic ray emissivity of the column behind the thermal gas can be measured from the excess of the observed brightness temperature of the HII region over its electron temperature (see \citealt{polderman2019} and references therein for a thorough treatment of this topic).  
Assuming the HII region G021.884$-$00.318 is resolved at our spatial resolution, the maximum depth of the absorption at 74~MHz, in brightness temperature, is $-$6500~K. This measurement is important for Galactic cosmic ray physics and can add to the growing catalogue of HII absorption regions \citep{polderman2020}. 
 
Figure~\ref{alpha-maps}d displays the spectral index distribution over Kes~69  obtained between the 74~MHz and 1.4~GHz maps from VLSSr and MAGPIS surveys respectively. There seems to be a general trend of flattening from northwest to southeast ($\alpha \sim -0.4$ to $-0.2$). This is compatible with the molecular shell in the vicinity of the remnant detected by 
\citet{zhou-09}.

The integrated spectral slope for Kes~69 indicates a low-frequency turnover (Fig.~\ref{74-spectra}d) suggestive of foreground thermal absorption, an issue we revisit in Sect.~\ref{ISM}.  The weighted fit to our compiled flux densities results in a radio spectral index $\alpha=-0.505\pm0.018$, in good agreement with the value $-0.5\pm0.11$ from K94, but somewhat discrepant with the S11 value of $\alpha=-0.56\pm0.03$ derived from a power-law fit to measurements at $\nu >$ 330~MHz. 
We note that part of the thermal emission from the HII region G021.884$-$00.318 could have been included in earlier, low-resolution flux density measurements of Kes~69 at frequencies higher than 74~MHz. However, since this HII region contributes only $\sim1$~Jy at GHz frequencies, its impact on the accuracy of the integrated SNR spectra is minimal. 

\vspace{0.5cm}
\noindent
\bf SNR W41 (G23.3$-$0.3): \rm 
The VLSSr image shown in Fig.~\ref{74-images}e clearly reveals the elongated  $\sim$18$^{\prime}$ western part of the SNR shell, with a highly irregular boundary and enhanced knots of emission at several locations. 
Radio continuum imaging at 1.4~GHz shows a weak arm emerging from the northern part of the western edge, extended about 20$^\prime$ to the east, which has been interpreted as a non-thermal component of the radio emission that originates from W41 \citep{lea08}. Traces of this northern W41 arm also appear in the GLEAM Survey, with a synthesised beam of 5$^{\prime}.6 \times 5^{\prime}.3$ at 88~MHz \citep{wayth15,hurley19}. However we do not see any corresponding weak emission in the VLSSr map above 4~$\sigma$ significance. Using this non-detection to set a limit on the spectral index for this emission, we estimate that this feature would contribute $\lesssim$10~\% to the integrated flux density of W41 at 74~MHz, which is well within the reported measurement errors.  

Figure~\ref{alpha-maps}e displays the spatial variations in the 74~MHz-1.4~GHz spectral index of the radio continuum emission created from VLSSr and MAGPIS images. The picture showing, on average, 
$\alpha \sim -0.35$ flat spectrum is consistent with the thermal gas 
in the region of the remnant. 
In Sect.~\ref{specials}, we further discuss the relative geometry of W41 and the ISM constituents near and towards it as discerned from its radio continuum spectrum.

The integrated spectral index from our spectrum in Fig.~\ref{74-spectra}e is $\alpha = -0.628\pm0.040$. Although our value is steeper than $\alpha=-0.48\pm0.14$ reported in K94, there is no significant differences because of the large uncertainty on their value. We are confident that our result represents a more reliable estimation, since our spectrum is better sampled, especially at frequencies lower than 200~MHz. 

Thermal sources of varying sizes ($0^\prime.7$ - $6^\prime.5$) within the area of this remnant are included in the WISE Catalog of Galactic HII regions \citep{and14}\footnote{An updated version of the \citet{and14}'s catalogue is available at \url{http://www.astro.phys.wvu.edu/wise/}}.  
These sources have been mapped in previously published radio continuum images of W41 at frequencies higher than 74~MHz (e.g. 330~MHz, \citealt{kassim-92}; 1.4~GHz, \citealt{lea08}).  Therefore, it is highly probable that contamination from this thermal gas has limited a precise estimate of the flux of the synchrotron emission from the remnant in that portion of the radio spectrum.

\vspace{0.5cm}
\noindent
\bf SNR Kes~73 (G27.4+0.0): \rm 
Originating $\lesssim2000$~yr ago from a core-collapse SN event, Kes~73 is one of the youngest SNRs in the Galaxy \citep{borkowski17}. The VLSSr image shows a slightly asymmetric shell with a bright, nearly point-like spot (Fig.~\ref{74-images}f). The position of this feature at R.A$\simeq$$18^{\mathrm{h}}\,41^{\mathrm~{m}}\,15^{\mathrm{s}}.6$, 
Decl.$\sim$$-04^{\circ}\,56^{\prime}\,59^{\prime\prime}$ does not coincide with the magnetar 1E~1841$-$045 thought to be the compact remnant of the stellar explosion that created G27.4+0.0 \citep[][and references therein]{kum14}.  From the current image it is not possible to tell if it is part of the SNR or an unrelated background source.  In the latter case, we note that it contains $\simeq$ 12\% of the total flux from Kes~73, well within the errors on the measured total flux.

The local distribution of the radio spectrum over the SNR, calculated between 74~MHz and 1.4~GHz from the comparison of VLSSr and MAGPIS images, is on average flatter than the integrated spectral index value  (Fig.~\ref{alpha-maps}f). It can be explained, however, in terms of the ionised material in the region of the remnant. A detailed consideration of the interstellar medium properties in connection with the radio emission from Kes~73 is presented in Sect.~\ref{specials}.

The inclusion of the 74~MHz flux density in the integrated continuum spectrum (Fig.~\ref{74-spectra}f) supports the low-frequency turnover inferred in \citet{kovalenko-94} and \citet{kas89s}. 
The fit to the flux measurements is overplotted in Fig.~\ref{74-spectra}f (although the 30.9~MHz upper limit appears in the spectrum it was excluded from the fit).   We find a spectral index $\alpha=-0.690\pm0.035$, in good agreement with the result in K94.

\vspace{0.5cm}
\noindent
\bf SNR G28.6$-$0.1: \rm 
This source has been poorly studied in the radio band since its identification as a Galactic SNR with a broken morphology \citep{helfand89}. The VLSSr map of G28.6$-$0.1 shows three bright connected structures (see Fig.~\ref{74-images}g). The absence of 74-MHz emission from nearby sources located at the northwest of the SNR is consistent with their thermal nature as first proposed by \citet{helfand89} and confirmed later by \citet{and11}.  The spectral index image in Fig.~\ref{alpha-maps}g constructed using 74-MHz VLSSr and 1.4~GHz MAGPIS data,  reveals a striking flattening to the southeast. The near-IR image of G28.6$-$0.1 presented by \citet{lee+19-FeII} shows some [FeII] thin filaments in this portion of the remnant, thought to be created by a radiative shock front moving through the ambient medium. 

Despite the paucity of literature flux density measurements, VLSSr, GLEAM, 330~MHz and 1.4~GHz measurements are adequate to constrain a fit to the continuum spectrum with a single power-law with index $\alpha = -0.690 \pm 0.057$ (see Fig.~\ref{74-spectra}g). To measure the impact of the ionised material on the integrated continuum spectrum of the source,  more sensitive observations below 100~MHz are needed.

\vspace{0.5cm}
\noindent
\bf SNR Kes~75 (G29.7$-$0.3): \rm 
Kes~75 is one of the strongest radio sources in our sample. The 74-MHz map shown in Fig.~\ref{74-images}h reveals an elliptically-shaped SNR with the brightest radio emission found on the southwestern part of the shell. There is no evidence for the PWN powered by the PSR~J1846$-$0258 \citep{got00} in the VLSSr image.  In contrast to most higher-frequency total-intensity radio images \citep[e.g. at 1.4  and 89~GHz in][]{boc05}, the full northern extent of the shell is visible in the VLSSr image.

The spectral index map of Kes~75 between the VLSSr at 74~MHz and MAGPIS at 1.4~GHz is displayed in Fig.~\ref{alpha-maps}h.  The southwest region of the remnant has an $\alpha~\sim~-0.4$, flatter than that of its surroundings by $\sim$+0.2. This spectral component corresponds to a location where the SN shock is running into a molecular shell and the brightness in radio continuum, X rays and mid-IR is strongest.

From the spectrum of Kes~75 plotted in Fig.~\ref{74-spectra}h, the 30.9~MHz flux density lies below the general trend of the data, suggesting a low frequency turnover (see also Sect.~\ref{ISM}). Excluding a turnover, the spectral index $\alpha = -0.659\pm0.014$ from our integrated spectrum is consistent with values found in both K94 and S11.

\vspace{0.5cm}
\noindent
\bf SNR 3C~391 (G31.9+0.0): \rm 
The VLSSr 74-MHz continuum image of 3C~391 presented in Fig.~\ref{74-images}i clearly shows a bright rim on the western side of the remnant, while towards the eastern half of the source the emission is dimmer. This picture is consistent with VLA imaging by \citet{bro05-391} at both 330~MHz and 74~MHz, in which they attribute localised absorption in this SNR to the interaction zone between the SNR and its immediate environment. They conclude that the thermal absorbing gas was created by the impact of a dissociative shock from the SNR with the molecular cloud with which it is interacting.  

The spectral index map for 3C~391 between the emission at 74~MHz from VLSSr and at 1.4~GHz from MAGPIS is displayed in Fig.~\ref{alpha-maps}i. The trend of spectral flattening is consistent with \citet{bro05-391}'s interpretation of a SNR interacting with a molecular cloud.   

The integrated spectrum of 3C~391 shows a robust low-frequency turnover (Fig.~\ref{74-spectra}i), which is seen in multiple low-frequency measurements. Our spectral index $\alpha = -0.521 \pm 0.004$ after combining the new VLSSr flux density measurements with previous published values (excluding the 30.9~MHz upper limit) is completely consistent with earlier results derived by \citet{bro05-391} and K94. We do not find evidence for two straight power laws with a frequency break at 1~GHz, which was presented in S11. We notice that the data points below 1~GHz in their spectrum are notoriously more scattered than ours.  We favour the low frequency turnover over the broken power law fit,  since the plotted measurements that meet our selection criteria show a clear smooth curve at low frequencies. 

\vspace{0.5cm}
\noindent
\bf SNR 3C~396 (G39.2$-$0.3): \rm 
The radio emission from this remnant has been almost exclusively mapped 
above 1~GHz \citep[][and references therein]{cru16}. To the best of our knowledge, the 74-VLSSr image (see Fig.~\ref{74-images}j) is the highest resolution sub-GHz view of this source presented to date. The source was imaged at 330 MHz by \citet{kassim-92} using the VLA, but was only marginally resolved.  Lower angular resolution images of 3C~396 from GLEAM at 88, 118, 155, and 188~MHz are available as well.   In the VLSSr map the SNR appears considerably distorted from circular symmetry. 

A bright ridge of emission extends from south to north, with a knot $\sim$1.4 times brighter than its surroundings. Between this position and a second radio enhancement further north, lies the non-thermal X-ray emission attributed to a central pulsar wind nebula powered by a putative pulsar \citep{olb03}.  As is also observed at higher frequencies, the radio emission of 3C~396 gradually fades from west to east.  A blowout tail towards the northeastern portion of the SNR shell and curving around to the west was visible in the radio domain using 1.4~GHz data \citep{patnaik-90}. It is also observed as an extended bright structure in the \it Spitzer \rm MIPSGAL image at 24~$\mu$m \citep{rea06}.  It is not visible in the 74~MHz VLSSr image, which supports the thermal emission mechanisms proposed to explain its origin, and is unlikely to be part of the SNR \citep{anderson93}. Furthermore, there is no structure in the VLSSr 74-MHz map which corresponds to the southwest extension noticeable in the earlier 330~MHz study of \citet{kassim-92}. We have the sensitivity to see this feature, but owing to its non-detection in the VLSSr image  we conclude it is either thermal or due to confusion. 

The spectral index map of 3C~396 between 74~MHz and 1.4~GHz using VLSSr and VGPS data is presented in Fig.~\ref{alpha-maps}j. Variations in the spectrum are recorded from about  $\alpha \sim -0.3$ up to $\sim~-0.6$. The flattest spectral feature occurs towards the southwestern corner of the remnant, a region where both [FeII] and H$_{2}$ near-IR line emission have been detected, indicating a SNR shock interacting with a dense medium \citep{lee+19-FeII}. 
Additional details of the thermal gas responsible for the 
spectral characteristics of 3C~396 are presented in 
Sect.~\ref{specials}. 

The integrated radio spectrum of 3C~396 is shown in  Fig.~\ref{74-spectra}j. The fit to the set of data with a power law and an exponential turnover model yields a spectral index $\alpha = - 0.351 \pm 0.010$. Our determination agrees well with the global spectral index reported in S11 ($\sim -0.34$) and \citet{cru16} ($\sim -0.364$). In these three cases the spectra include high frequency fluxes up to $\sim$33~GHz. 
We also notice that our result (and hence those from S11 and \citealt{cru16}) differs from the global spectral index derived by K94 ($\sim -0.48$). We believe the spectrum presented in K94 is less reliable because the lack of measurements at frequencies higher than 10.6~GHz.  

\vspace{0.5cm}
\noindent
\bf SNR 3C~397 (G41.1$-$0.3): \rm 
As seen in Fig.~\ref{74-images}k, the  morphology of this SNR at 74~MHz follows the box-like shape observed at higher radio frequencies \citep[e.g.][]{dyer-reynolds-99}. The brightest region is on the west side of the source and contains $\sim30$\% of the total flux density measured at 74~MHz (see Table~\ref{74properties}).  Publicly available images for this source at 1.4GHz do not include the full flux density reported in the literature, so we have not constructed a spectral index map for it.

As noticed by \citet{dyer-reynolds-99}, confusion with thermal emission from an HII region just west of the SNR (since catalogued as G041.126$-$00.232, \citealt{and14}) likely prevented accurate non-thermal measurements of the SNR by many lower resolution instruments which are reported in the literature.  The spectral index $\alpha=-0.356\pm0.013$ that we have fit (Fig.~\ref{74-spectra}k) is hardly compatible with the $-0.46 \pm 0.10$ value reported in K94, but it is considerably flatter than the spectrum  $\alpha=-0.50\pm0.01$ measured by S11 using a pure power-law fit to a much less complete set of flux density measurements. 

\vspace{0.5cm}
\noindent
\bf SNR W49B (G43.3$-$0.2): \rm 
The VLSSr 74-MHz image in Fig.~\ref{74-images}l shows a roughly circular source, approximately $5^\prime$.5 in diameter, with considerably brightened emission on the eastern part of the shell.  
The spectral map created from the 74~MHz VLSSr and 1.4~GHz MAGPIS images is shown in Fig.~\ref{alpha-maps}k. There is a dramatic flattening from the northeast ($\alpha \sim~-0.8$) to the southwest portions of W49B ($\alpha \sim~-0.15$). 
Overall the distribution of spectral index shows the same trends as in \citet{lac01}.

A spectral turnover for $\nu \lesssim 100$~MHz in the integrated spectrum is well known and has been linked by \citet{lac01} to foreground thermal gas superimposed over the western half of the remnant.  The low-frequency thermal absorption is very distinctive in the new spectrum shown in Fig.~\ref{74-spectra}l, and the derived value of the global spectral index $\alpha =-0.461\pm0.009$ is consistent with the spectral fit by \citet{lac01}. 
In Sect.~\ref{specials} we readdress the analysis of the thermal gas responsible for the absorption implied by the radio spectrum of W49B.

\vspace{0.5cm}
\noindent
\bf Tycho's SNR (G120.1+1.4): \rm 
The VLSSr low-frequency image of this rim-brightened shell-type SNR agrees with previously published images at higher radio frequencies \citep[e.g.][]{katz-stone-00}. Figure~\ref{74-images}m displays a roughly circular shell of $\sim$9$^{\prime}$ in diameter with a highly non-uniform emission. There is some departure from circular symmetry in the southeast. A brightening is especially prominent in the northeastern portion of the SNR where the peak surface brightness is $\sim$14~Jy~beam$^{-1}$.  This enhancement could be produced by the northeast front of the shell impinging the inner boundary of the wind-blown molecular bubble, the latter revealed in $^{12}$CO $J=2-1$ line observations \citep{zho16}. As with all classic rim-brightened shell-type SNRs, the interior emission is much more diffuse. Overall, the surface brightness distribution in the VLSSr image of Tycho is also in reasonably good agreement with that observed with LOFAR in the 58-143~MHz range \citep{arias+19}. Fig.~\ref{alpha-maps}l displays the spatial spectral variations across Tycho~SNR, obtained from VLSSr and NVAS data at 74~MHz and 1.4~GHz, respectively. Our spectral index map nicely picks out and confirms the internal absorption observed by \citet{arias+19} with the LOw Frequency ARray (LOFAR).

The integrated radio spectrum is shown in Fig.~\ref{74-spectra}m, based on data collected over more than three decades from $\sim$15~MHz to 70~GHz. It yields a relatively steep integrated spectrum $\alpha=-0.624\pm0.004$, largely consistent with the previous determination reported by K94 but incompatible with that of S11 ($-0.58 \pm 0.02$). Alternatively, \citet{reynolds92} modelled the emission from Tycho at frequencies up to $\sim$10~GHz with a modest spectral break in the power law of $\Delta\alpha\simeq0.04$ at 1~GHz, inferring an underlying curved spectrum. Power-law fits to our data below and above the inferred break yield slopes of 
$\alpha_{<1~\mathrm{GHz}} = -0.607\pm0.012$ and 
$\alpha_{>1~\mathrm{GHz}} = -0.581\pm0.010$, 
respectively. This is weakly consistent with an intrinsically curved, underlying concave-up spectrum.  

The steep radio spectral index and suggested curvature in this remnant have been attributed by several authors to non-linear effects in the acceleration process of the radio-emitting electrons (e.g. \citealt{reynolds92}, \citealt{volk2008}).
\citet{wilhelm2020-astho} recently presented an alternative explanation, suggesting the entire spectrum, from radio to $\gamma$-rays, can be reproduced in terms of stochastic re-acceleration in the immediate downstream region of the SNR forward shock, without invoking the consequences of non-linear particle acceleration kinetic theory.

\vspace{0.5cm}
\noindent 
\bf SNR 3C~58 (G130.7+3.1): \rm 
3C~58 represents the archetypal example of a pulsar-powered plerion. 

In the 74~MHz image (Fig.~\ref{74-images}n), it appears elongated in the east-west direction, with a size of approximately $8^{\prime}.5\times4^{\prime}.5$, similar to that observed at higher radio frequencies and even in X rays \citep{sla04,bietenholz-06}. 

Due to the limited spatial resolution, the VLSSr image does not resolve the complex of loop-like features seen throughout the nebula at higher radio frequencies, but shows a bright component at the location of the pulsar PSR~J0205+6449, and a brightness distribution that gradually fades with radial distance from the centre.  The morphology of the spectral index between 74~MHz and 1.4~GHz from the VLSSr and NVAS images, respectively, is shown in Fig.~\ref{alpha-maps}m. It matches the flat and uniform 74/327~MHz and 327~MHz/4.9~GHz spectral index distributions presented by \citet{bietenholz-01}.  

The integrated spectrum is presented in Fig.~\ref{74-spectra}n and includes measurements at frequencies from 38~MHz to 217~GHz, with a notable gap between $\sim 5$ and $\sim$14~GHz. The best-fit to the data points from 38~MHz to 5~GHz results in a synchrotron spectral index $\alpha=-0.076\pm0.008$, in agreement with previous studies (see, for instance, \citealt{green-92}, \citealt{kovalenko-94} or \citealt{sun11}), while measurements above 14~GHz are better fit by a considerably steeper power-law index $\alpha = -0.383\pm0.022$. 
From our analysis, a spectral break occurs at $\sim$12~GHz, somewhat lower than the break reported by \citet{planck2016}. However, within the uncertainties, our result is compatible with the value  $\sim$18~GHz provided by \citet{xu+19}, which results from the fit to a broadband (from radio to X rays)  spectrum of  3C~58. As in the case of  G21.5$-$0.9,  an interplay of synchrotron cooling and  re-acceleration of electrons in the pulsar wind nebula through the ASA process was used by  \citet{xu+19} to explain the observed spectral shape. Adding new radio measurements in the 5-14~GHz gap is critical to refining the spectral break in 3C~58.

\section{Analysis of the Low Frequency Spectral Turnovers} 
\label{ISM}

\subsection{General Considerations}

Turnovers in the low frequency ($\nu\lesssim100$~MHz) continuum spectra of SNRs were first identified in the 1970’s and attributed to external thermal absorption by an ionised component of the ISM along the line of sight \citep[see e.g.][]{dul75}. With a larger sample, \citet{kas89s} attributed the observed patchiness of the absorption to low-density, intervening extended HII region envelopes (EHEs), the existence of which had been inferred earlier from stimulated, meter-wavelength radio recombination lines \citep{ana86}. In the 1990's intrinsic SNR thermal absorption was first detected in Cas~A due to unshocked ejecta \citep{kas95}, from thermal filaments in the Crab Nebula \citep{bie97}, and much more recently in Tycho with LOFAR \citep{arias+19}. A third possible source of thermal absorption is ionised gas resulting from physical processes in the immediate surroundings of SNRs. All three cases are important because they provide critical distance constraints for disentangling the relative superposition of ionised gas and SNRs.

One of the earliest detections of resolved thermal absorption towards a Galactic SNR was made at the relatively high frequency of 330 MHz in the special case of the Galactic centre, through the detection of Sgr A West against the Sgr A East SNR \citep{anantha91}. Thereafter followed W49B, the earliest typical case of resolved foreground absorption, attributed to an EHE enveloping a complex of HII regions along the line of sight \citep{lac01}. By contrast and also within the VLSSr sample, 3C~391 represents the first detection of resolved thermal absorption at the interface of a SNR blast wave interacting with a molecular cloud\footnote{We also refer the reader to the study presented by \citet{cas11} on IC~443.} \citep{bro05-391}. 

In this section we use our spectral fits of the eight SNRs in our sample exhibiting turnovers to constrain the properties of the absorbing thermal gas.

\subsection{Absorption by Intervening Ionised Gas}

In four cases (Kes~67, Kes~69, Kes~75, and 3C~397, see Table~\ref{table2}), we can easily attribute the moderately low optical depths and higher levels of absorption at lower frequencies to the generic case of foreground ionised gas which is not associated with the SNRs.   For simplicity we assume that EHEs are the most likely source of this gas, but we note that other manifestations of intervening ionised gas could also cause the absorption. 

To estimate the physical properties of the intervening ionised gas we can make use of the relation between the  optical depth at a reference  frequency, $\tau_{\nu_0}$, the electron temperature of the thermal gas $T_\mathrm{{e}}$, and 
the  emission measure 
$\mathrm{EM}=\displaystyle{\int_{L}}n_{\mathrm{e}}^2\, dx$, which depends in turn on  the electron density, $n_{\mathrm{e}}$, and the  linear extent $L$ along the line of sight. Setting the singly-ionised species, $Z_\mathrm{i} = n_\mathrm{e}/n_\mathrm{i} = 1$, we have the expression  \citep{wilson09} 

\begin{equation}\label{EM}
\tau_{\nu_{0}}= 3.014 \times 10^{-2} \left(\frac{T_{\mathrm{e}}}{\mathrm{K}}\right)^{-3/2}\left(\frac{\nu_{0}}{\mathrm{GHz}}\right)^{-2} \, g_{\mathrm{ff}} \,\left(\frac{\mathrm{EM}}{\mathrm{pc~cm^{-6}}}\right),   
\end{equation}

\noindent
where 
$g_\mathrm{ff}$ is the Gaunt factor: 

\begin{equation}\label{gaunt}
g_{\mathrm{ff}}= \mathrm{ln}\left[4.955 \times 10^{-2}\left(\frac{\nu_{0}}{\mathrm{GHz}}\right)^{-1}\right] + 1.5\, \mathrm{ln}\left(\frac{T_{\mathrm{e}}}{\mathrm{K}}\right).   
\end{equation}

For the four spectral turnover cases we attribute to line of sight absorption, we adopt generic EHE electron temperatures and path lengths of 5000~K and 100~pc, respectively \citep{kas89s}, inferring the EM and electron densities reported in Table~\ref{table2}.  These are consistent with properties inferred independently from RRLs free of discrete continuum sources \citep{anantha85b}, as expected given that the generic EHE properties from \citet{kas89s} were also consistent with the RRL results.

\subsection{Absorption by associated ionised gas}
\label{specials}
Interpretation of the remaining SNRs with low-frequency spectral turnovers is less clear. EHE absorption seems questionable to account for the relatively high optical estimate for W41, while infrared (IR) and molecular line emission from Kes~73, 3C~396 and W49B show intriguing correspondence with their non-thermal radio emission. The SNR 3C~391 also shows evidence of correlation between IR and non-thermal radio emission. Because it has been studied in depth by \citet{bro05-391}, and our new spectra are consistent with their results, we do not discuss it further here.

For SNRs Kes~73 and W49B, we construct free-free optical depth maps using the relation $\tau_{\mathrm{74}}=\ln\left(S_\mathrm{exp}/S_\mathrm{obs}\right)$, where $S_\mathrm{obs}$ is the observed 74~MHz emission, while $S_\mathrm{exp}$ is the emission expected if no absorption is present.  To create the $S_\mathrm{exp}$ maps we scale the 1.4~GHz images from the literature to the expected 74 MHz fluxes using the integrated radio spectral indices of each source (Kes~73, $\alpha\simeq-0.69$ and W49B, $\alpha\simeq-0.46$, see Table~\ref{table2}). The extrapolated images are convolved to match the 75$^{\prime\prime}$ resolution of the VLSSr images, and masked at the 4-$\sigma$ level based on their respective noise. Based on reasonable assumptions for the electron temperatures, we convert the $\tau_\mathrm{74}$ distributions into local EM measurements based on Eq.~\ref{EM}. An important assumption in the extrapolation from 1.4~GHz to 74~MHz is that measured spectral deviations are dominated by foreground thermal absorption and not by intrinsic variations in the spectrum of the synchrotron emitting electrons, which are typically much more subtle.

\subsubsection{SNR~W41} \label{W41}
\begin{figure}[ht!]
  \centering
\includegraphics[width=0.4\textwidth]{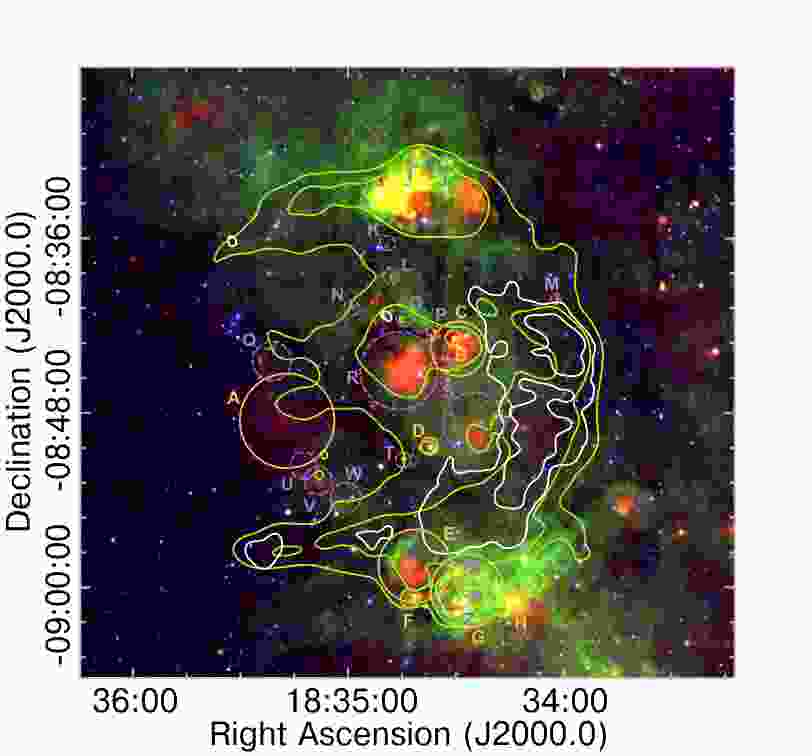}  
\caption{\it Spitzer \rm 3-colour image (RGB: 24, 8.0, and 3.6~$\mu$m) towards the SNR~W41's complex with contours of radio continuum emission overlaid. Green contours (at 0.25 and 0.42~Jy~beam$^{-1}$ levels) correspond to the 1.4~GHz image from MAGPIS, convolved to match the VLSSr resolution of $75^{\prime\prime}$. 
VLSSr 74~MHz continuum contours are superimposed in white (levels: 0.56 and 1.30~Jy~beam$^{-1}$). Multiple HII regions, A-H, in the field  with known distances near W41 ($\sim$5~kpc) are marked by orange circles, while thermal sources (I-Z) with unknown distance are indicated by dashed violet circles \citep{and14}.}
\label{W41fig}
\end{figure}

Figure~\ref{W41fig} shows the mid-IR emission from GLIMPSE and the MIPS Galactic Plane Survey \citep{carey09} in color with superimposed radio emission contours for SNR~W41. There are numerous HII regions with line-of-sight coincidence with the W41 SNR, many of which overlap the 1.4 GHz emission (green contours). The remnant and probably the majority of these HII regions are part of the giant molecular cloud G23.0$-$0.4 (\citealt{hogge19}, and references therein). For HII regions A-H (see Fig.~\ref{W41fig}) the association with the SNR is readily accepted because their previously established distances (e.g. from radio recombination lines, \citealt{chen20}) are consistent with that of the remnant ($4.8 \pm 0.2$~kpc, \citealt{rana18}). No distance measurements have been reported for the remaining HII regions I-Z. 

Compared to the emission at 1.4~GHz, the 74~MHz VLSSr emission is significantly attenuated on the eastern side of the remnant (white contours in Fig.~\ref{W41fig}) and bears striking testimony to thermal absorption. \footnote{Note that the HII regions are not expected in emission at 74 MHz, a frequency at which they are almost certainly optically thick; if anything they would appear in absorption against the Galactic background or the SNR, but the limited resolution and sensitivity of the 74 MHz map preclude direct detection.} This is consistent with the prevalence of HII regions on the eastern side of the complex. While this indicates that many of these HII regions (and any associated EHEs) must be in the foreground relative to the SNR, this need not be the case for all of them. For example HII regions E and M could be background to the SNR, since they appear almost coincident with regions where the 74 MHz emission is not attenuated. Nevertheless, on the basis of the VLSSr image we are able to place an independent upper limit $\sim$5~kpc to a significant subset of these HII regions, based on the SNR distance. Future observations are required to precisely place all the HII regions relative to the SNR, but it would not be surprising if they are all roughly at the same distance and associated with the G23.0$-$0.4 giant molecular cloud complex. 

To estimate the physical characteristics of the ionised gas responsible for absorbing the 74~MHz~SNR emission, we first derive the electron temperature by using the relation between magnitude and the Galactocentric distance to the source (\citealt{quireza06}, more detail given in the analysis of G21.8$-$0.6~SNR in Sect.~\ref{individual}). For the complex of HII regions in which  W41 is embedded at an assumed $\sim$5~kpc heliocentric distance, we find a characteristic  $T_{\mathrm{e}}\simeq7100$~K. Under the high-opacity approximation, valid for low-frequency radio measurements, we calculate the average brightness temperature of the synchrotron emission behind the thermal gas of $T_{\mathrm{B}}\sim6000$~K. Future RRL observations are required to better constrain the electron temperatures, but we note that our rough estimate is consistent with those measured in typical HII regions \citep{luisi2019}. 

Assuming that the ionised gas can be characterised by a single electron temperature, we can derive the average EM corresponding to the absorbing gas from Eq.~\ref{EM}, using the measured optical depth from our best-fit model to the SNR spectrum ($\tau_\mathrm{74}\sim1.2$). 
We derived a mean $\mathrm{EM}\simeq10 \times 10^{3}$~pc~cm$^{-6}$, comparable to the values typically measured in other known HII regions \citep{luisi2019}. This result can be used to estimate the electron density of the absorbing gas using the observed geometry. If we consider a path-length $L\sim4$~pc through the ionised gas, equivalent to the mean linear size of the overlapping HII regions (angular diameters ranges from $\sim$$1^\prime.5$ to $\sim$$7^\prime.5$), then the resulting average electron density is $n_\mathrm{e} \simeq 60$~cm$^{-3}$ (Table~\ref{table2}). This density likely represents a blend of absorption from denser HII region cores and their associated lower density EHEs.

\subsubsection{SNR~Kes~73}\label{Kes73}
\begin{figure*}[ht!]
  \centering
 \includegraphics[width=0.9\textwidth]{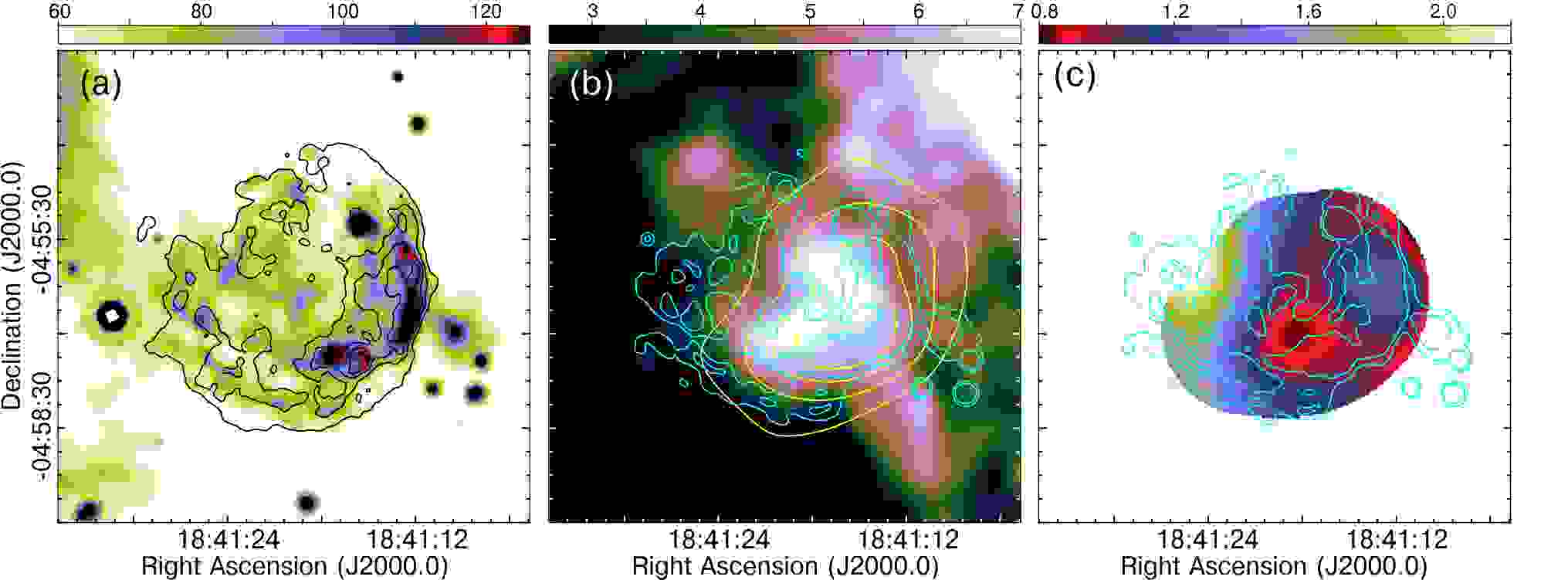}       
\caption{\it Panel a: \rm \it Spitzer \rm MIPSGAL 24~$\mu$m image for SNR~Kes~73 (shown with a linear colour scale in MJy~sr$^{-1}$) overlaid with black contours tracing 1.4~GHz radio continuum emission at levels 2.5, 4.8, 7.6, and 11~mJy~beam$^{-1}$ from the MAGPIS survey. 
The morphology of the IR emission strongly mimics the radio emission.
\it Panel b: \rm  $^{13}$CO (1-0) data from GRS integrated  in the $v_{\mathrm{LSR}}$=95-105~km~s$^{-1}$ range. 
The linear colour scale is in K~km~s$^{-1}$. 
Cyan and yellow contours represent the mid-IR (levels: 70 and 84~MJy~sr$^{-1}$) and 74~MHz low-radio frequency intensities (levels: 0.8, 1.3, 1.8, and 2.2~Jy~beam$^{-1}$), respectively. The newly-identified CO cloud is detected within the boundaries of the Kes~73 shell. 
\it Panel c: \rm Optical depth towards Kes~73 at 74~MHz as a function of position. Cyan contours superimposed delineate the  mid-IR emission as in panel \it b\rm.}
\label{Kes73new}
\end{figure*}

Kes~73 is very bright at 24~$\mu$m, and this emission is completely coincident with the SNR radio shell as shown in Fig.~\ref{Kes73new}a. Taking the properties of the observed mid-IR features into consideration, \citet{carey09}  argued it could predominantly arise from [OIV] and [FeII] line emission. Later, \citet{pinheiro2011} attributed the IR radiation in Kes~73 to dust grains heated by collisions in the hot plasma behind the SNR shock front.  The emission is analogous to other known SNR molecular cloud interactions in which ionised atomic species created after the passage of a dissociating SNR shock produce abundant line emission in the mid-IR (e.g. RCW~103, W44, W28, 3C~391, \citealt{oliva1999, reach2000}).

The striking coincidence of the mid-IR emission with the 1.4 GHz radio emission is consistent with the low frequency spectral turnover depicted in Fig.~\ref{74-spectra}f. We hypothesise that the turnover can be plausibly attributed to absorption by ionised gas in a molecular cloud that is either in the process of being enveloped by the SNR shock wave or has already been impacted by it. This interpretation is supported by near-IR [FeII]-emitting ($\sim$1.6~$\mu$m) clumps detected in the southern part of Kes~73; these were interpreted to be shocked circumstellar gas rather than high-speed metal-enriched SN ejecta \citep{lee+19-FeII}.

The scenario of a SNR-molecular cloud interaction is supported by Fig.~\ref{Kes73new}b, which depicts the integrated intensity map from the Boston University-FCRAO Galactic Ring Survey (GRS, \citealt{jac+06}) in the velocity range $v_{\mathrm{LSR}}$= 95-105~km~s$^{-1}$ overlaid with the mid-IR (cyan contours) and 74~MHz radio continuum (yellow contours) emission. The plotted molecular emission includes the velocity corresponding to the $\sim$5.8~kpc kinematic distance to Kes~73 recently revised by \citet{rana18} and \citet{lee2020}.  A bright portion of the CO cloud is spatially coincident with the SNR, with molecular emission also extending to the northwest of Kes~73.  Around the peak of the cloud we find features with a velocity width of $\sim$8-10~km~s$^{-1}$.  Most of the mid-IR gas shows good spatial correlation with the molecular gas, especially towards the west. 
It is noteworthy that the molecular structure mapped here is not consistent with the previous study by \citet{liu17}, who considered molecular gas  at a distance of 9~kpc associated with  the western ($\sim$90~km~s$^{-1}$) and northwestern ($\sim$85-88~km~s$^{-1}$) boundaries of the remnant.

To test the reliability of our idea that the observed absorption is tracing ionised gas in the cloud interacting with the SNR, we examined the characteristics of the absorption by locally mapping the free-free continuum optical depth at 74~MHz across Kes~73. To do this, we used the 74-MHz VLSSr and the 1.4~GHz-MAGPIS image of the remnant. 

As shown in Fig.~\ref{Kes73new}c the optical depth varies by a factor of $\sim 3$ across the source, with average errors in $\mathrm{\tau_{74}}$ of $~$25\%. Mid-IR emission contours from \it Spitzer \rm at 24~$\mu$m are superposed on the $\tau_{74}$ optical depth map, generally indicating IR emission widely distributed across the high optical depth features. The mid-IR displays a saddle shape, with peaks to the east and west, and a minimum near the centre. The optical depth map mimics this morphology, exactly as expected if the region of highest mid-IR emission corresponds to the region of strongest low frequency absorption (a similar effect was seen in 3C~391 by \citealt{bro05-391}). Note that the IR contours lying outside of the eastern boundary of the optical map correspond to a very low surface brightness region at 74~MHz, which was clipped for the construction of the optical depth map. 

Typical electron density can be derived from the distribution of $\mathrm{EM}$ (not shown here) assuming, as is usual in the literature, that absorption occurs over a characteristic mean path length equivalent to the mean transverse extent of the region where the optical depth changes.    We use this assumption to characterise the approximate range of electron densities in the ionised gas corresponding to the non-uniform absorption inferred in Fig.~\ref{Kes73new}c. 
We measure absorption levels corresponding to characteristically high and low optical depths of 1.8 and 0.95, respectively. Furthermore we adopt $T_{\mathrm{e}}\sim7000$~K from the mid-IR ionic line observations of interstellar shocks \citep{hewitt+09-IR}. Following Eq.~\ref{EM} and Eq.~\ref{gaunt} we derive local variations in $\mathrm{EM}\simeq$ 8-15 $\times 10^{3}$~pc~cm$^{-6}$. Considering mean size scales ranging between 
$L\simeq$ 0.$^{\prime}$7 and 1$^{\prime}$ (or $\sim$1.2-1.7~pc at a distance of $\sim$5.8~kpc to Kes~73) for the optical depth variations, we obtain thermal absorbing electron densities $n_{\mathrm{e}}$=$\sqrt{\mathrm{EM}/L}\sim$ 70-110~cm$^{-3}$ for the ionised gas causing the absorption towards Kes~73. 
While these values are much larger than that estimated in the extended envelopes of HII regions (0.5-10~cm$^{-3}$, \citealt{kas89s}), they are consistent with electron densities derived from low-radio frequency data in the ionised shocked gas linked to 3C~391, an established SNR molecular cloud interaction \citep{bro05-391}. 
Our estimate is similarly consistent with the electron density reported in \citet{koo2016} ($\sim$600~cm$^{-3}$) from the analysis of mid-IR [FeII] line ratios in the shocked gas of a sample of SNRs.
A partial explanation for the difference between the electron density computed here and that by \citet{koo2016}
might be that their value corresponds to a model with a shock speed  150~km~s$^{-1}$, which is slower than expected for a young SNR with a relatively fast shock speed such as Kes~73
(age $\sim$1400~yr, blast wave velocity $\sim 1400 \,d_{8.5}$~km~s$^{-1}$ $\simeq$ 950~km~s$^{-1}$, after correction by the revised 5.8~kpc SNR's distance, \citealt{borkowski17}). 
We note that significant variations in electron densities $\sim$100-1000~cm$^{-3}$ can be inferred from combinations of different shock models or mid-IR ionic lines ratios with modest temperature variations typically centred at $\sim$7000~K \citep{oliva1999,reach2000, hewitt+09-IR,koo2016}.

We conclude that the ionised gas associated with the interaction of Kes~73 and a molecular cloud are likely to be responsible for the observed low frequency radio absorption. The absorbing gas may be ionised directly by the interaction of the shock front with the cloud, or from the SNR's X-ray radiation \citep{kum14}. Since Kes~73 is a relatively young SNR, ionization by its stellar progenitor cannot be discounted either. Finally, it is not unreasonable to consider absorption by unshocked ejecta such as is found in Cas~A or Tycho SNRs \citep{arias+18,arias+19}. However based on a spatially resolved spectroscopic X-ray study \citep{kum14} and the non-centrally condensed, widespread distribution of high optical depth across the remnant, we find this latter scenario unlikely.

\subsubsection{W49B}
One of the earliest examples of spatially resolved thermal absorption against a Galactic SNR was made at 74~MHz  by \citet{lac01}. They attributed the significant attenuation towards the southwest region of the remnant to foreground absorption by intervening HII regions and their associated EHEs. The resolved absorption was consistent with the low frequency turnover in the integrated SNR spectrum (for example see Fig. ~\ref{74-spectra}l). The inference of intervening HII regions seemed consistent with much earlier detection of radio recombination lines by \citet{pankonin76} towards W49B's non-thermal radio shell. However much higher resolution, modern observations (c.f. \citealt{and14}) reveal no classic HII regions towards the western part of W49B, where \citet{lac01} measured the strongest absorption. 
The only catalogued source is the foreground HII region WISE~G043.305$-$00.211 (2$^\prime$ in size) located at 4.3~kpc \citep{and14} towards the northeastern edge of W49B. Furthermore, \citet{kalcheva18} identified the thermal WISE source as an ultracompact HII region  (named G043.3064$-$00.2114) with a size of 2$^{\prime\prime}$ at a distance of 4.4~kpc. Given the poor angular resolution ($\sim$8$^{\prime}.5$, two times the SNRs' size) of \citet{pankonin76}'s observations it is probable their RRL detection emanated from the thermal source G043.305$-$00.211. 

\begin{figure*}[ht!]
  \centering
\includegraphics[width=0.7\textwidth]{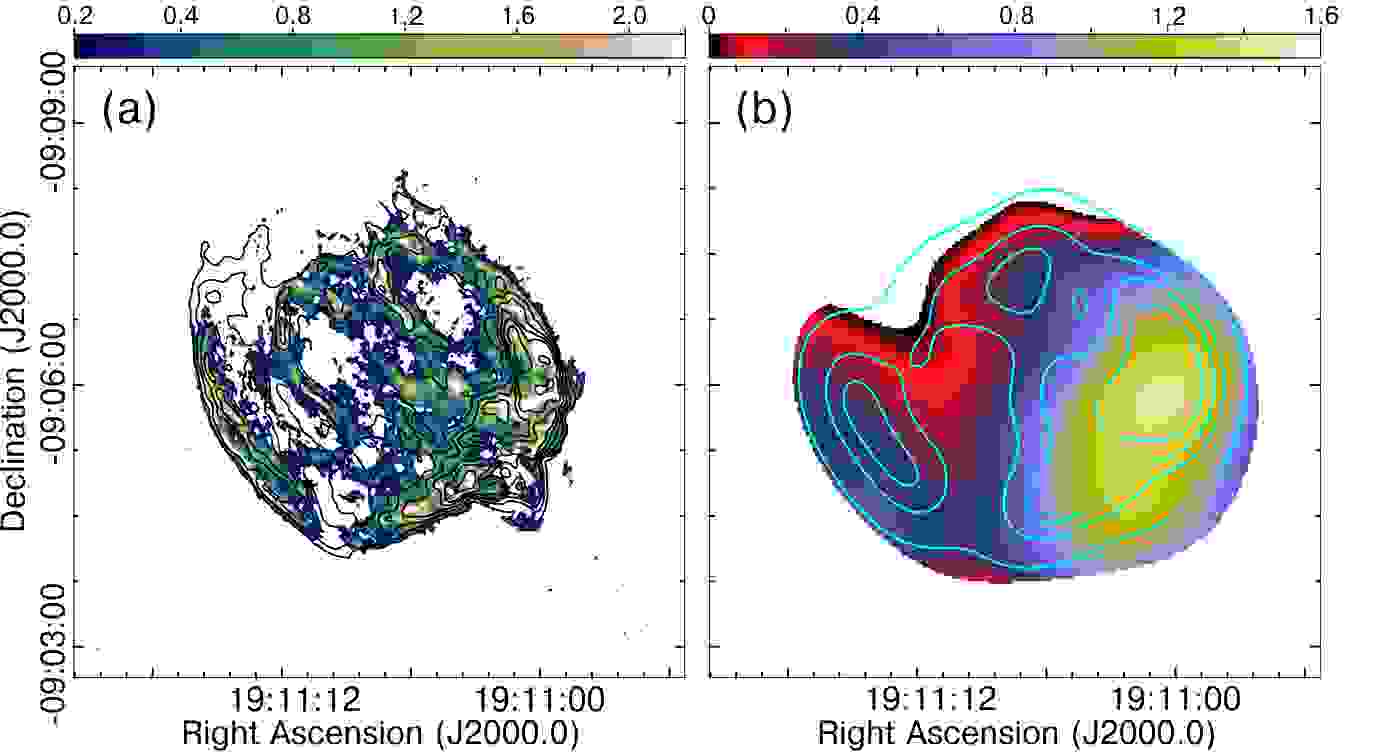}
  \caption{
 \it Panel a: \rm [FeII] line emission at 1.644~$\mu$m towards SNR~W49B (image kindly provided by Dr. Lee, Y.-H.)  overlaid with 
 contours (levels: 3, 10, 20, 30, 40, 50, and 60~mJy~beam$^{-1}$) of the 1.4~GHz continuum radiation detected in the MAGPIS survey. The colour representation is linear in units of MJy~sr$^{-1}$. 
 \it Panel b: \rm Local distribution of the optical depth computed towards W49B at 74~MHz. Cyan contours of the near-IR [FeII] line emission matched at the 45$^{\prime\prime}$ resolution are superimposed for comparison, with levels at 12, 47, 70, and 105~MJy~sr$^{-1}$.
}
\label{W49B}
\end{figure*}

Infrared observations, presented by \citet{keohane+07} and more recently by \citet{lee+19-FeII}, have shown that W49B is very bright in both [FeII] (1.644~$\mu$m) and $\mathrm{H_{2}}$ (2.122~$\mu$m) line emission.  There is good spatial agreement between filaments emitting in ionic iron lines and the synchrotron radio shell of the SNR (see Fig.~\ref{W49B}a), while the $\mathrm{H_{2}}$ near-IR emitting gas encloses the eastern, southern, and western SNR boundaries.  There are several scenarios to explain the apparent association of [FeII] emission with W49B.  They include shock interaction with material swept up by the stellar winds of the progenitor, radiative atomic shocks propagating into a dense ambient medium, 
photoionisation by the adjacent X-ray emission from the shock-heated ejecta, and SN ejecta with high Fe abundance (\citealt{lee+19-FeII}, and references therein). 
While the current evidence is not sufficient to  exclude X-ray heating, the $\mathrm{H_{2}}$ (2.122~$\mu$m) emission favors a shock interaction with dense material.   In addition, previous studies of the large-scale ambient medium of W49B have shown molecular clouds close to the remnant \citep{simon01, zhu+14}.

The considerably improved observations towards W49B since \citet{lac01} suggest a significantly revised and more exciting interpretation of the low frequency observations than the chance superposition of HII regions along the line of sight. 
Figure~\ref{W49B}b shows the distribution of the free-free optical depth in W49B constructed from the radio continuum emission towards the source at 74~MHz  and 1.4~GHz from the VLSSr and NVSS surveys, respectively. Superposed contours trace the [FeII] near-IR line emission.  The absorption levels across the SNR are high, ranging from 0.2 to 1.6.  The highest absorption features are localised in the eastern and western portions of W49B (mean $\tau_{74}\sim$ 0.45 and 1.4, respectively) and show excellent correspondence with the brightest [FeII] filaments despite the limited angular resolution of the $\tau_{74}$ map. This correspondence provides strong evidence that \textit{the thermal absorption in W49B derives from a direct interaction with its environment}.

In order to estimate the  thickness of the ionised gas layer, we analyse variations in the $\mathrm{EM}$ with position over W49B from the local values of the optical depth at 74~MHz. We adopt an electron temperature of 10$^{4}$~K, with an electron density of $\sim$500~cm$^{-3}$ (estimated from the ionic gas in W49B by \citealt{zhu+14}). 
Using  Eq.~\ref{EM} and Eq.~\ref{gaunt}, the average free-free optical depth towards the eastern and western portion of the SNR shell translate into $\mathrm{EM}$ values 
between 6 $\times$ 10$^{3}$ and 19 $\times$ 10$^{3}$~pc~cm$^{-6}$, respectively. 
From this we conclude that the observed absorption is taking place in a thin layer with typical thickness of  $L=\mathrm{EM}/n_\mathrm{e}^2$=(2.5-7.6)$\times$10$^{-2}$~pc ($\sim$0.8-2.3 $\times$10$^{17}$~cm). These values, markedly  narrower than the typical thickness of $\sim$5~pc measured in ISM cavities surrounding SNRs \citep{fukui12}, are consistent with the hypothesis that absorption of the low frequency radio emission is due to ionised material generated by the impact of the SN shock front on the wall of the bubble shaped by the winds of W49B's progenitor star \citep{keohane+07}.

\subsubsection{3C~396} 
Spectroscopic observations made with \it Spitzer \rm reveal multiple ionic (e.g., [FeII], [NeII], [SiII], etc.) and molecular (e.g. $\mathrm{H_{2}S(0)}$, $\mathrm{H_{2}S(1)}$, etc.) transitions across SNR 3C~396 \citep{hewitt+09-IR}. 
A very bright [FeII] ($\sim$1.6~$\mu$m) filament is observed near the southwestern boundary of the remnant along with $\mathrm{H_{2}}$ ($\sim$2.1~$\mu$m) emission extending outside of the radio and the [FeII]-line boundaries (\citealt{lee+19-FeII}, and references therein). The ionic lines are indicative of a post-shock colliding region. The shocked ambient medium could be either dense clumps previously formed by the wind material of the SN progenitor \citep{lee09-3C396} or a dense molecular cloud interacting with the 3C~396 blast wave \citep{lee+19-FeII}. \citet{su11} claim that $^{12}$CO (1-0 and 2-1) molecular material at the distance of 6.2~kpc is colliding with the SNR shock front. Subsequently, \citet{rana18} revised the distance to the source to $8.5\pm0.5$~kpc, based on HI 21~cm and $^{13}$CO line observations.  This is roughly consistent with the $9.5\pm0.1$~kpc estimate derived by \citet{lee2020} from the velocities of the near-IR H$_{2}$ emission lines. Thus the interaction of 3C~396 with the ionic line emission and CO at 6.2~kpc is uncertain.

\begin{figure}
 \centering
 \includegraphics[width=0.45\textwidth]{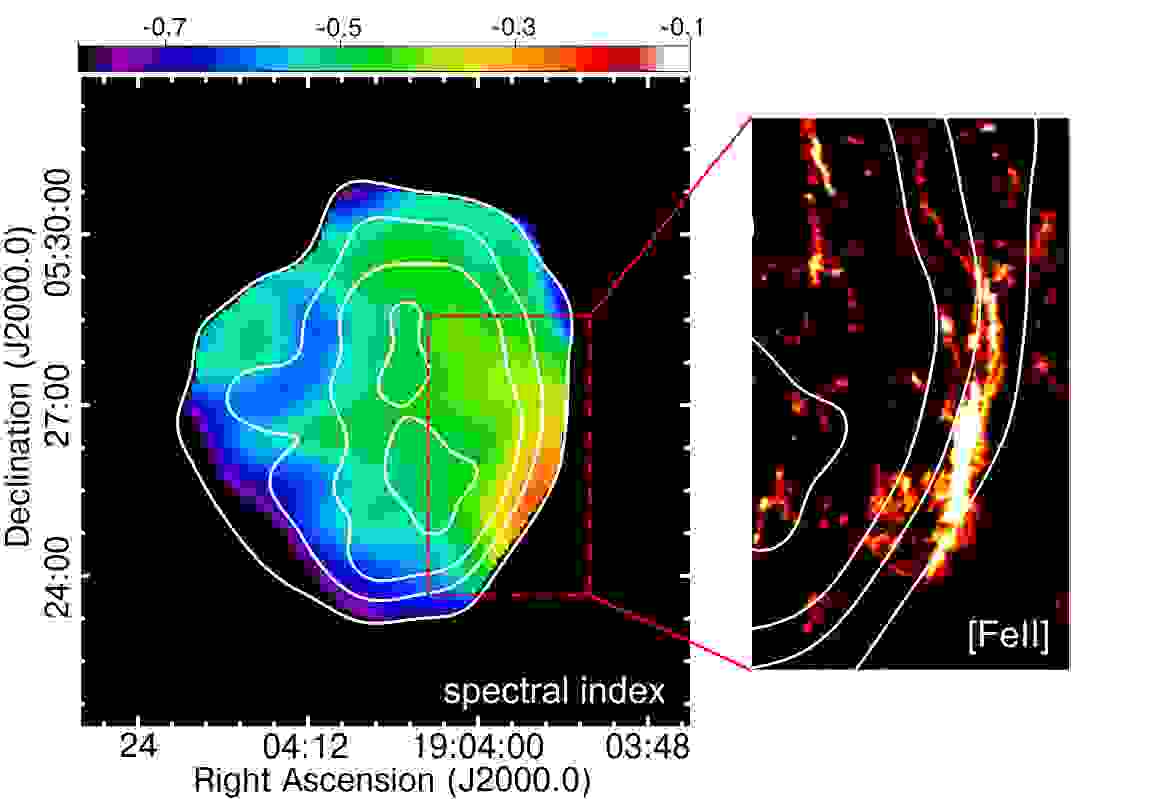}
 \caption{Comparison between the 74~MHz/1.4~GHz radio spectral index distribution over the remnant 3C~396 (the same displayed in Fig.~\ref{alpha-maps}) and the filaments emitting in ionic iron line emission. The flattest spectral index feature is located where the bright [FeII] near-IR filaments are observed. Contours from the 74-MHz VLSSr image at levels 0.64, 1.7, 2.5, and 3.6~Jy~beam$^{-1}$ are included for reference.}
\label{special-3C396}
\end{figure}

The spectral index map for this source (Fig.~\ref{alpha-maps}j) reveals variations on the order of $\pm$ 0.35 across 3C~396, from $-$0.26 in the south west, gradually steepening to $\alpha \sim$ $-$0.6 towards the interior of the western shell. The spectral flattening is discernible in an extension of $\sim$0.8$^{\prime}$ $\times$ 2$^{\prime}$.5 along the southwestern limb, with spectral index $-$0.26 to $-$0.35. A striking result enabled by our spatially resolved spectral index map is the correspondence between this flattest region and the bright [FeII] filament noted above, as seen in Figure ~\ref{special-3C396}. This immediately implies thermal absorption in the post-shock interaction region traced by the ionic line emission, reminiscent of 3C~391. It is not surprising that the integrated spectrum for this source eventually turns over at lower frequencies as the effects of the absorption grow stronger (Fig. ~\ref{74-spectra}i). These results show that this bright SNR, along with 3C~391, would be an excellent target for higher resolution, lower frequency observations to perform a comprehensive analysis of the correlation between the ionised IR gas and thermal absorption discerned from the radio optical depth. 

\section{Summary and Conclusions}
\label{summary}
We have updated the radio continuum spectra of 14 Galactic SNRs using new flux density measurements from the VLSSr and GLEAM in combination with carefully selected measurements from the literature.  Where possible, measurements over the range of frequencies from  50~MHz to 50~GHz were placed on the absolute flux density scale developed by \citet{per17}. The spectra were fit by power-laws, broken power-laws, and power laws with low frequency turnovers as appropriate for each source. We have measured steep spectral index values ($|\alpha|>0.5$) for the younger non-PWNe sources in our sample. 

The VLSSr data independently confirm the area of absorption in the centre of the Tycho SNR seen by LOFAR LBA \citep{arias+19}, although the integrated spectrum is well-fit by a simple power-law without a turnover.   The integrated spectra for SNR G4.5+6.8 (Kepler) and G28.6$-$0.1 are also well fit by simple power-laws, and no evidence for thermal absorption is seen.    The integrated spectra for the two pulsar wind nebulae, G21.5$-$0.9 and 3C~58 (G130.7+3.1), are well fit by power-laws with spectral breaks at 38 and 12~GHz, respectively.  The new integrated spectrum for SNR 3C~391 confirms the low-frequency turnover previously seen by \citet{bro05-391}.  

We analyse potential free-free thermal absorption processes that cause the turnovers at frequencies below 100~MHz observed in the spectra of eight additional SNRs. We explain the curved spectra of Kes~67, Kes~69, Kes~75, and 3C~397 as thermal absorption occurring when their lines of sight intersect thermal gas, which may be diffuse ionised envelopes associated with normal HII regions. 

For SNR~W41, we explain the thermal absorption as arising in a number of HII regions either in close proximity to, or coincident with, the SNR. The average extension of the non-thermal radio emission in this remnant at 74 MHz is $20\%$ smaller than the combined thermal and non-thermal emission measured at 1.4~GHz. This indicates that the size of the SNR had been previously overestimated due to contamination by thermal sources. 

For the three SNRs Kes~73, 3C~396, and W49B,  we have found a strong spatial correspondence between the IR ionic line emission and the highest level of absorption that we measure in these sources.  On the basis of this correlation and previously reported evidence for the interaction of these sources with their surroundings, we explain the free-free absorption towards these sources in terms of ionised gas created after the impact of the SN shock with the interstellar material. For Kes~73 and W49B we are able to derive electron temperatures and densities. This is similar to the interpretation of 3C~391 previously published by \citet{bro05-391}.  Our study adds to a growing body of work demonstrating that physical questions about SNRs and their surroundings can be tackled by incorporating low-radio frequency observations in the analysis, since these data are a potent tool for separating thermal and non-thermal emission in complex regions.

\section{Future Work}

The improved accuracy of the integrated continuum spectra achieved by adding reliable measurements below 100~MHz is important to better understand the physical processes within SNRs. For example, the improved integrated spectra can help constrain theories that explain high-energy particle production in a sample of SNRs which are also bright in $\gamma$ rays (G21.5$-$0.9, W41, Kes~73,  Kes~75, 3C~391, W49B, Tycho, and 3C~58). A future work, analysing the broad-band spectral distribution of these sources is in preparation and it will be presented elsewhere for publication.

Despite the progress represented in this paper, the paucity of good measurements below 100~MHz, where thermal absorption is much easier to detect, remains prevalent. There are many SNRs, e.g. Kes~75, where the inference of a low-frequency turnover relies on a single low frequency measurement. Emerging data from the LOFAR LBA Sky Survey (LoLSS:  \citealt{lolss+21}) survey, with unprecedented resolution and sensitivity below 100~MHz, should have a major impact on this field. Despite LOFAR's limited access to the inner Galaxy and the majority of known SNRs, it should significantly increase the known population of interacting SNRs detectable through thermal absorption.

\begin{acknowledgements}
We wish to acknowledge the comments from the anonymous referee. 
G. Castelletti and L. Supan are members of the {\it Ca\-rre\-ra del Investigador Cient\'{\i}fico} of CONICET, Argentina. 
This research was partially supported by the grant awarded by the ANPCyT (Argentina) PICP~2017-3320. 
Basic Research at the Naval Research Laboratory is funded by 6.1 base programs. 
This publication makes use of molecular line data from the Boston University-FCRAO Galactic Ring Survey (GRS). 
The authors want to thank Dr. Yong-Hyun Lee for kindly providing the [FeII] emission images used in this work.
\end{acknowledgements}

\bibliographystyle{aa}
 \bibliography{castelletti-supan-21}
 
\appendix
\section{List of integrated flux densities}\label{all_flux_values}
Tables in this section contain  the flux density measurements used in this work to model the radio continuum spectra of 14 Galactic SNRs. As indicated in their third columns, the listed values include new low frequency fluxes that we calculated from the VLSSr and GLEAM projects, along with measurements from previous publications. 
We applied correction factors to adjust the fluxes between 50~MHz and 50~GHz to the absolute flux density scale of \citet{per17}, the frequency range over which the scale is valid with an accuracy of 3\%-5\%.
We notice that, out of the 454 flux densities listed in our work, only about 6\% have not been placed in the \citet{per17} scale because they correspond to either frequencies lower than 50~MHz or higher than 50~GHz. 
Additionally, for a small fraction (14\%) of the fluxes a determination of the correction factor was not possible because information on the primary flux calibrators was not  recorded in the original reference. 

%---------------------------------------------------------------------------------- G4.5+6.8
\begin{table}[h!]
\small
\centering
\caption{Integrated flux densities compiled for the SNR G4.5+6.8, used to trace the spectrum presented in Fig.~\ref{74-spectra}a.}
\label{flux_densities_list_G4.5}
%----------------------------------------------------------------------------------
\begin{tabular}{c l l}\hline\hline\\[-5pt]
\multicolumn{3}{c}{\bf G4.5+6.8 (Kepler)}                                           \\[4pt]\hline
%----------------------------------------------------------------------------------
 Frequency       &   Scaled      flux         &  \multirow{2}{*}{Reference}   \\
 $[$MHz$]$       &  density       [Jy]         &                               \\\hline
     74	&    111 $\pm$ 17	    &	This work (from VLSSr)	               \\
     80	&	 98.4 $\pm$	9.6	 &	\citet{dulk-slee-72}	   \\
     80	&  100.9 $\pm$	30.3   &	\citet{slee77}	           \\
     80	&   93.6 $\pm$	7.6	 &	\citet{dulk-slee-75}	   \\
     83	&	104.4 $\pm$	14.9   &	\citet{kovalenko-94}	   \\
     86 	&	110.0 $\pm$	22.0\tablefootmark{~\dag} &	\citet{mills-slee-hill-60} \\
     88 	&	100.0 $\pm$	9.0	 &	This work (from GLEAM)	       \\
     102	&	 86.8 $\pm$	12.8   &	\citet{kovalenko-94}       \\
     111	&	 84.7 $\pm$	12.8   &	\citet{kovalenko-94}	   \\
     118	&	 77.0 $\pm$	7.0	 &	This work (from GLEAM)	       \\
     155	&	 62.0 $\pm$	9.0	 &	This work (from GLEAM)	       \\
     160	&	 55.0 $\pm$	16.5   &	\citet{slee77}             \\
     408	&	 42.6 $\pm$	6.4	 &	\citet{dulk-slee-72}       \\
     635	&	 26.2 $\pm$	2.6	 &	\citet{milne-hill-69}	   \\
     960	&	 21.2 $\pm$	2.9	 &	\citet{harris-62}	       \\
     960	&	 18.8 $\pm$	2.0	 &	\citet{harris-roberts-60}  \\
     960	&	 21.2 $\pm$	0.5	 &	\citet{trushkin-98}	       \\
    1000	&	 20.0 $\pm$	3.0\tablefootmark{~\dag}	 &	\citet{milne-70}           \\
    1410	&	 16.7 $\pm$	1.7	 &	\citet{milne-hill-69}      \\
    1420	&	 13.2 $\pm$	1.6	 &	\citet{lequeux-62}	       \\
    2300	&	 12.2 $\pm$	0.3	 &	\citet{trushkin-98}	       \\
    2650	&	 12.3 $\pm$	1.2	 &	\citet{milne-hill-69}	   \\
    2700	&	  9.1 $\pm$	0.9	 &	\citet{milne-dickel-74}	   \\
    3900	&	  8.0 $\pm$	0.2	 &	\citet{trushkin-98}	       \\
    4850	&	  6.4 $\pm$	0.3	 &	\citet{griffith-94}	       \\
    5000	&	  6.6 $\pm$	1.3	 &	\citet{milne-dickel-75}    \\
    5000	&	  7.2 $\pm$	1.1	 &	\citet{dulk-slee-72}       \\\hline

\end{tabular}
\tablefoot{
\tablefoottext{\dag}{No correction factor to bring the original flux density measurement to the scale of \citet{per17} was applied because no information on the flux density calibrators was found in the original publication. }
}
\end{table}
%----------------------------------------------------------------------------------

%---------------------------------------------------------------------------------- G18.8+0.3
\begin{table}[h!]
\small
\centering
\caption{Integrated flux densities compiled for the SNR G18.8+0.3, used to trace the spectrum presented in Fig.~\ref{74-spectra}b.}
\label{flux_densities_list_G18.8}
%---------------------------------------------------------------------------------
\begin{tabular}{c l l}\hline\hline\\[-5pt]
\multicolumn{3}{c}{\bf G18.8+0.3 (Kes~67)}                                         \\[4pt]\hline
%---------------------------------------------------------------------------------
  Frequency   &     Scaled    flux        &  \multirow{2}{*}{Reference} \\
  $[$MHz$]$   &  density        [Jy]        &                             \\\hline
 30.9	&	87.4	$\pm$	17.5\tablefootmark{~*}	&	\citet{kassim-88}	\\
 57.5	&	73.9	$\pm$	 8.2	&	\citet{odegard-86}	\\
  74	&	76.2	$\pm$	13.8	&	This work (from VLSSr)	\\
  80	&	84.1	$\pm$	11.8	&	\citet{slee-higgins-73}	\\
  80	&	66.9	$\pm$	19.1	&	\citet{dulk-slee-72}	\\
  83	&	86.5	$\pm$	19.9	&	\citet{kovalenko-94}	\\
  88	&	  81	$\pm$	  17	&	This work (from GLEAM)	\\
 111	&	78.8	$\pm$	19.7	&	\citet{kovalenko-94}	\\
 118	&	  80	$\pm$	  14	&	This work (from GLEAM)	\\
 155	&	  71	$\pm$	  11	&	This work (from GLEAM)	\\
 160	&	67.7	$\pm$	10.1	&	\citet{dulk-slee-75}	\\
 200	&	  62	$\pm$	   9	&	This work (from GLEAM)	\\
 330	&	50.8	$\pm$	 7.6	&	\citet{kassim-92}	\\
 408	&	38.9	$\pm$	 5.8	&	\citet{kesteven-68}	\\
 408	&	49.2	$\pm$	 7.4	&	\citet{kassim-89-list}	\\
 960	&	28.3	$\pm$	 0.3	&	\citet{trushkin-98}	\\
1000	&	35.0	$\pm$	 5.3\tablefootmark{~\dag}	&	\citet{milne-70}	\\
1465	&	29.7	$\pm$	 0.3	&	\citet{dubner-96}	\\
2695	&	18.9	$\pm$	 1.9	&	\citet{altenhoff-70}	\\
2700	&	22.3	$\pm$	 3.3	&	\citet{milne-dickel-74}	\\
3650	&	17.4	$\pm$	 0.3	&	\citet{trushkin-98}	\\
3900	&	18.3	$\pm$	 0.3	&	\citet{trushkin-98}	\\
4800	&	15.1	$\pm$	 0.9	&	\citet{sun11}	\\
5000	&	17.9	$\pm$	 3.6	&	\citet{milne-dickel-75}	\\
5000	&	14.6	$\pm$	 2.2	&	\citet{milne-69}	\\
5000	&	13.1	$\pm$	 2.6	&	\citet{reifenstein-70}	\\
8400	&	12.9	$\pm$	 1.0\tablefootmark{~\dag}	&	\citet{milne-89}	\\
\hline
\end{tabular}
\tablefoot{
\tablefoottext{\dag}{No correction factor to bring the original flux density measurement to the scale of \citet{per17} was applied because no information on the flux density calibrators was found in the original publication. } \\
\tablefoottext{*}{No correction to \citet{per17} was applied because the measurement is out of the 50~MHz-50~GHz frequency range where their scale is valid.  }
}
\end{table}
%---------------------------------------------------------------------------------

%-------------------------------------------------------------------------------------------- G21.5-0.9
\begin{table}[h!]
\small
\centering
\caption{ Integrated flux densities compiled for the SNR G21.5$-$0.9, used to trace the spectrum presented in Fig.~\ref{74-spectra}c.}
\label{flux_densities_list_G21.5}
%--------------------------------------------------------------------------------------------
\begin{tabular}{c l l}\hline\hline\\[-5pt]
\multicolumn{3}{c}{\bf G21.5$-$0.9}                                                  \\[4pt]\hline
%--------------------------------------------------------------------------------------------
 Frequency   &     Scaled    flux           &  \multirow{2}{*}{Reference} \\
 $[$MHz$]$   &  density        [Jy]           &                             \\\hline
  30.9	&  10.4	$\pm$	2.1\tablefootmark{~*}		&	\citet{kassim-88}	\\
  57.5	&	6.2	$\pm$	1.2	&	\citet{kassim-89-list}	\\
    74	&	6.4	$\pm$	1.1	&	This work (from VLSSr)	\\
    83	&	8.9	$\pm$	2.0	&	\citet{kovalenko-94}	\\
    88	&	7.2	$\pm$	2.7	&	This work (from GLEAM)	\\
   111	&	7.9	$\pm$	2.0	&	\citet{kovalenko-94}	\\
   118	&	6.6	$\pm$	2.0	&	This work (from GLEAM)	\\
   155	&	6.3	$\pm$	1.3	&	This work (from GLEAM)	\\
   160	&	7.7	$\pm$	2.3	&	\citet{slee77}	\\
   200	&	5.9	$\pm$	1.1	&	This work (GLEAM)	\\
   330	&	8.5	$\pm$	1.3	&	\citet{kassim-92}	\\
  1400	&	6.7	$\pm$	1.0\tablefootmark{~\dag}		&	\citet{frail-moffett-93}	\\
  1660	&	6.5	$\pm$	1.3\tablefootmark{~\dag}		&	\citet{milne-69}	\\
  2300	&	5.9	$\pm$	0.3	&	\citet{trushkin-98}	\\
  2700	&	5.7	$\pm$	1.1\tablefootmark{~\dag}		&	\citet{milne-69}	\\
  2700	&	6.9	$\pm$	0.7	&	\citet{milne-dickel-74}	\\
  2700	&	7.0	$\pm$	0.4	&	\citet{becker-kundu-76}	\\
  2700	&	6.8	$\pm$	1.4	&	\citet{goss-day-70}	\\
  3900	&	6.2	$\pm$	0.3	&	\citet{trushkin-98}	\\
  4800	&	6.4	$\pm$	0.4	&	\citet{sun11}	\\
  4875	&	6.9	$\pm$	1.0	&	\citet{downes-80}	\\
  5000	&	6.7	$\pm$	1.3	&	\citet{reifenstein-70}	\\
  5000	&	6.7	$\pm$	0.3\tablefootmark{~\dag}		&	\citet{bietenholz-bartel-08} \\
  5000	&	6.1	$\pm$	1.8	&	\citet{altenhoff-70}	\\
  7700	&	6.3	$\pm$	0.3	&	\citet{trushkin-98}	\\
  8100	&	7.0	$\pm$	0.4\tablefootmark{~\dag}		&	\citet{becker-kundu-76}	\\
  8900	&	6.2	$\pm$	0.9	&	\citet{caswell-clark-75}	 \\
 10600	&	6.4	$\pm$	0.7	&	\citet{becker-kundu-76}	     \\
 11200	&	6.0	$\pm$	0.3	&	\citet{trushkin-98}	         \\
 32000  &   5.0 $\pm$   0.3 &   \citet{morsi-reich-87}       \\
 70000  &   4.3 $\pm$   0.6\tablefootmark{~*}	 &   \citet{planck2016}           \\
 84000  &   3.9 $\pm$   0.7\tablefootmark{~*}	 &   \citet{salter-89b}           \\
 90700  &   3.8 $\pm$   0.4\tablefootmark{~*}	 &   \citet{salter-89a}           \\
 94000  &   3.5 $\pm$   0.4\tablefootmark{~*}	 &   \citet{bock-01}              \\
100000  &   2.7 $\pm$   0.5\tablefootmark{~*}	 &   \citet{planck2016}           \\
143000  &   3.0 $\pm$   0.4\tablefootmark{~*}	 &   \citet{planck2016}           \\\\\hline
\end{tabular}
\tablefoot{
\tablefoottext{\dag}{No correction factor to bring the original flux density measurement to the scale of \citet{per17} was applied because no information on the flux density calibrators was found in the original publication.}\\
\tablefoottext{*}{No correction to \citet{per17} was applied because the measurement is out of the 50~MHz-50~GHz frequency range where their scale is valid. }}
\end{table}
%--------------------------------------------------------------------------------------------

%---------------------------------------------------------------------------- G21.8-0.6
\begin{table}[h!]
\small
\caption{Integrated flux densities compiled for the SNR G21.8$-$0.6, used to trace the spectrum presented in Fig.~\ref{74-spectra}d.}
\label{flux_densities_list_G21.8}
%----------------------------------------------------------------------------
\begin{tabular}{c l l}\hline\hline\\[-5pt]
\multicolumn{3}{c}{\bf G21.8$-$0.6 (Kes~69)}                                    \\[4pt]\hline
%----------------------------------------------------------------------------
 Frequency  &   Scaled      flux              &  \multirow{2}{*}{Reference} \\
 $[$MHz$]$  &  density       [Jy]              &                             \\\hline
29.9	&	210 $\pm$ 40\tablefootmark{~*}	&  \citet{jones74} 	\\
30.9	&	170.6  $\pm$	34.1\tablefootmark{~*}	&	\citet{kassim-88}	\\
57.5	&	172.4  $\pm$	34.5	&	\citet{kassim-89-list}	\\
  74	&	169.0  $\pm$	31.0	&	This work (from VLSSr)	\\
  83	&	208.8  $\pm$	39.8	&	\citet{kovalenko-94}	\\
  88	&	194	   $\pm$	21	    &	This work (from GLEAM)	\\
 111	&	177.2  $\pm$	39.4	&	\citet{kovalenko-94}	\\
 118	&	166	   $\pm$	20	    &	This work (from GLEAM)	\\
 155	&	143    $\pm$	16	    &	This work (from GLEAM)	\\
 200	&	116    $\pm$	14	    &	This work (from GLEAM)	\\
 330	&	121.4  $\pm$	18.2	&	\citet{kassim-92}	\\
 408	&	73.0   $\pm$	7.3	    &	\citet{shaver-goss-70b}	\\
 960	&	63.7   $\pm$	6.1	    &  	\citet{trushkin-98}	\\
1000	&	64.0   $\pm$	9.6\tablefootmark{~\dag}		&	\citet{milne-70}	\\
1414	&	47.9   $\pm$	4.8 	&	\citet{altenhoff-70}	\\
1660	&	41.5   $\pm$	6.2\tablefootmark{~\dag}		&	\citet{milne-69}	\\
2700	&	41.0   $\pm$	8.2\tablefootmark{~\dag}		&	\citet{milne-69}	\\
2700	&	31.7   $\pm$	9.2 	&	\citet{willis-73}	\\
2700	&	42.1   $\pm$	4.6 	&	\citet{velusamy-kundu-74}	\\
2700	&	39.6   $\pm$	7.9 	&	\citet{milne-dickel-74}	\\
2700	&	38.5   $\pm$	7.7 	&	\citet{goss-day-70}	\\
3650	&	29.1   $\pm$	2.9 	&	\citet{trushkin-98}	\\
3900	&	27.0   $\pm$	2.9 	&	\citet{trushkin-98}	\\
4800	&	23.7   $\pm$	1.3 	&	\citet{sun11}	\\
4875	&	26.7   $\pm$	4.0 	&	\citet{downes-80}	\\
5000	&	23.1   $\pm$	4.6 	&	\citet{kundu-74}	\\
5000	&	29.0   $\pm$	5.8 	&	\citet{milne-dickel-75}	\\
5000	&	27.1   $\pm$	5.4 	&	\citet{goss-shaver-70}	\\
\hline
\end{tabular}
\tablefoot{
\tablefoottext{\dag}{No correction factor to bring the original flux density measurement to the scale of \citet{per17} was applied because no information on the flux density calibrators was found in the original publication.}\\
\tablefoottext{*}{No correction to \citet{per17} was applied because the measurement is out of the 50~MHz-50~GHz frequency range where their scale is valid. }}
\end{table}
%----------------------------------------------------------------------------

%-------------------------------------------------------------------- G23.3-0.3
\begin{table}[h!]
\small
\centering
\caption{Integrated flux densities compiled for the SNR G23.3$-$0.3, used to trace the spectrum presented in Fig.~\ref{74-spectra}e.}
\label{flux_densities_list_G23.3}
%--------------------------------------------------------------------
\begin{tabular}{c l l}\hline\hline\\[-5pt]
\multicolumn{3}{c}{\bf G23.3$-$0.3 (W41)}                            \\[4pt]\hline
%--------------------------------------------------------------------
 Frequency &    Scaled     flux           &  \multirow{2}{*}{Reference} \\
 $[$MHz$]$ &    density     [Jy]           &                             \\\hline
 57.5	&	59.2	$\pm$	11.8	&	\citet{kassim-89-list}	\\
   74	&	88.0	$\pm$	17.0	&	This work (from VLSSr)	\\
   83	&	133.2	$\pm$	29.8	&	\citet{kovalenko-94}	\\
   88	&	154	    $\pm$	30  	&	This work (from GLEAM)	\\
  111 	&	143.7	$\pm$	29.5	&	\citet{kovalenko-94}	\\
  118	&	172 	$\pm$	40	    &	This work (from GLEAM)	\\
  155	&	149	    $\pm$	37	    &	This work (from GLEAM)	\\
  200	&	139 	$\pm$	28	    &	This work (from GLEAM)	\\
  330	&	138.7	$\pm$	27.7	&	\citet{kassim-92}	\\
  400	&	130.0	$\pm$	26.0\tablefootmark{~\dag}		&	\citet{downes-71}	\\
  408	&	148.6	$\pm$	22.3	&	\citet{kesteven-68}	\\
 1400	&	50.0	$\pm$	10.0\tablefootmark{~\dag}		&	\citet{downes-71}	\\
 1414	&	52.1	$\pm$	5.2 	&	\citet{altenhoff-70}	\\
 1420	&	59.7	$\pm$	8.2\tablefootmark{~\dag}	 	&	\citet{tian-07}	\\
 2695	&	35.9	$\pm$	3.6 	&	\citet{altenhoff-70}	\\
 5000	&	24.6	$\pm$	2.5 	&	\citet{altenhoff-70}	\\
 5000	&	24.0	$\pm$	5.0\tablefootmark{~\dag}	 	&	\citet{downes-71}	\\\hline
\end{tabular}
\tablefoot{
\tablefoottext{\dag}
{No correction factor to bring the original flux density measurement to the scale of \citet{per17} was applied because no information on the flux density calibrators was found in the original publication.}\\
%{The correction factor to bring the original flux density measurement to the scale of \citet{per17} was not available.} 
}
\end{table}
%--------------------------------------------------------------------

%------------------------------------------------------------------------ G27.4+0.0
\begin{table}[h!]
\small
\centering
\caption{Integrated flux densities compiled for the SNR G27.4+0.0, used to trace the spectrum presented in Fig.~\ref{74-spectra}f. }
\label{flux_densities_list_G27.4}
%------------------------------------------------------------------------
\begin{tabular}{c l l}\hline\hline\\[-5pt]
\multicolumn{3}{c}{\bf G27.4+0.0 (Kes~73)}                        \\[4pt]\hline
%------------------------------------------------------------------------
 Frequency     &   Scaled flux       &  \multirow{2}{*}{Reference} \\
 $[$MHz$]$     &    density     [Jy]       &                             \\\hline
   74	&	13.8	$\pm$	2.5	&	This work (from VLSSr)	\\
   80	&	18.2	$\pm$	5.5	&	\citet{slee77}	\\
 85.7	&	20.0	$\pm$	6.0\tablefootmark{~\dag}	&	\citet{mills-58}	\\
   88	&	17.0	$\pm$	3.5	&	This work (from GLEAM)	\\
  118	&	17.6	$\pm$	2.9	&	This work (from GLEAM)	\\
  155	&	14.6	$\pm$	3.0	&	This work (from GLEAM)	\\
  160	&	20.9	$\pm$	6.3	&	\citet{slee77}	\\
  408	&	10.4	$\pm$	1.6\tablefootmark{~\dag}	&	\citet{caswell-82}	\\
  408	&	13.0	$\pm$	2.0	&	\citet{kassim-89-list}	\\
  408	&	12.6	$\pm$	1.9	&	\citet{kesteven-68}	\\
 1415	&	 3.5	$\pm$	0.5\tablefootmark{~\dag}		&	\citet{caswell-82}	\\
 4850	&	 2.1	$\pm$	0.1	&	\citet{griffith-94}	\\
 5000	&	 1.9	$\pm$	0.2\tablefootmark{~\dag}		&	\citet{haynes-78}	\\
 5000	&	 1.9	$\pm$	0.5	&	\citet{angerhofer-77}	\\
 5000	&	 1.4	$\pm$	0.2	&	\citet{milne-69}	\\\hline
\end{tabular}
\tablefoot{
\tablefoottext{\dag}
{No correction factor to bring the original flux density measurement to the scale of \citet{per17} was applied because no information on the flux density calibrators was found in the original publication.}}
\end{table}
%------------------------------------------------------------------------

%------------------------------------------------------------------- G28.6-0.1
\begin{table}[h!]
\small
\centering
\caption{Integrated flux densities compiled for the SNR G28.6$-$0.1, used to trace the spectrum presented in Fig.~\ref{74-spectra}g.}
\label{flux_densities_list_G28.6}
%-------------------------------------------------------------------
\begin{tabular}{c l l}\hline\hline\\[-5pt]
\multicolumn{3}{c}{\bf G28.6$-$0.1}                               \\[4pt]\hline
%-------------------------------------------------------------------
  Frequency  &      Scaled  flux         & \multirow{2}{*}{Reference} \\
  $[$MHz$]$  &  density      [Jy]         &                            \\\hline
   74	&	26.9	$\pm$	4.7	&	This work (from VLSSr)	\\
   88	&	29.5	$\pm$	5.6	&	This work (from GLEAM)	\\
  118	&	25.9	$\pm$	3.9	&	This work (from GLEAM)	\\
  155	&	20.2	$\pm$	3.1	&	This work (from GLEAM)	\\
  200	&	15.5	$\pm$	1.9	&	This work (from GLEAM)	\\
  330	&	 9.9	$\pm$	0.8\tablefootmark{~\dag}		&	\citet{supan-12}	\\
 1420	&	 4.2	$\pm$	0.5\tablefootmark{~\dag}		&	\citet{supan-12}	\\\hline
\end{tabular}
\tablefoot{
\tablefoottext{\dag}
{No correction factor to bring the original flux density measurement to the scale of \citet{per17} was applied because no information on the flux density calibrators was found in the original publication.}}
\end{table}
%-------------------------------------------------------------------

%------------------------------------------------------------------------------ G29.7-0.3
\begin{table}[h!]
\small
\centering
\caption{Integrated flux densities compiled for the SNR G29.7$-$0.3, used to trace the spectrum presented in Fig. \ref{74-spectra}h.}
\label{flux_densities_list_G29.7}
%------------------------------------------------------------------------------
\begin{tabular}{c l l}\hline\hline\\[-5pt]
\multicolumn{3}{c}{\bf G29.7$-$0.3 (Kes~75)}                             \\[4pt]\hline
%------------------------------------------------------------------------------
 Frequency  &     Scaled    flux        & \multirow{2}{*}{Reference} \\
 $[$MHz$]$  &  density       [Jy]        &                         \\\hline
  30.9	&	31.2	$\pm$	6.2\tablefootmark{~*}		&	\citet{kassim-88}	\\
    74	&	48.5	$\pm$	7.9	&	This work (from VLSSr)	\\
    80	&	35.3	$\pm$	4.8	&	\citet{dulk-slee-75-to}	\\
    83	&	36.8	$\pm$	7.0	&	\citet{kovalenko-94}	\\
   102	&	38.5	$\pm$	10.8	&	\citet{kovalenko-94}	\\
   111	&	40.4	$\pm$	11.8	&	\citet{kovalenko-94}	\\
   118	&	42.3	$\pm$	6.5	&	This work (from GLEAM)	\\
   155	&	32.4	$\pm$	3.4	&	This work (from GLEAM)	\\
   160	&	26.9	$\pm$	6.7	&	\citet{slee77}	\\
   160	&	31.9	$\pm$	5.2	&	\citet{dulk-slee-75-to}	\\
   200	&	25.5	$\pm$	2.7	&	This work (from GLEAM)	\\
   408	&	19.0	$\pm$	1.9	&	\citet{shaver-goss-70a}	\\
   408	&	14.0	$\pm$	1.4	&	\citet{shaver-goss-70b}	\\
   408	&	18.5	$\pm$	2.8	&	\citet{green-74}	\\
   408	&	15.0	$\pm$	2.2	&	\citet{dulk-slee-72} \\
   408	&	15.0	$\pm$	2.3	&	\citet{kassim-89-list} \\
   960	&	9.9 	$\pm$	0.9	&	\citet{trushkin-98}	\\
  1400	&	7.0 	$\pm$	1.4\tablefootmark{~\dag}		&	\citet{downes-71}	\\
  1414	&	7.3 	$\pm$	2.2	&	\citet{altenhoff-70}	\\
  2695	&	5.0 	$\pm$	1.5	&	\citet{altenhoff-70}	\\
  2700	&	5.3 	$\pm$	0.5	&	\citet{milne-dickel-74}	\\
  4800	&	3.6 	$\pm$	0.6	&	\citet{sun11}	\\
  4875	&	2.8 	$\pm$	0.3	&	\citet{downes-80}	\\
  5000	&	3.4 	$\pm$	0.3	&	\citet{milne-69}	\\
  5000	&	3.2 	$\pm$	0.5	&	\citet{downes-71}	\\
  5000	&	3.4 	$\pm$	0.5	&	\citet{dulk-slee-72}	\\
 32000	&	0.9 	$\pm$	0.1	&	\citet{morsi-reich-87}	\\\hline
\end{tabular}
\tablefoot{
\tablefoottext{\dag}
{No correction factor to bring the original flux density measurement to the scale of \citet{per17} was applied because no information on the flux density calibrators was found in the original publication.}\\
\tablefoottext{*}{No correction to \citet{per17} was applied because the measurement is out of the 50~MHz-50~GHz frequency range where their scale is valid. }}
\end{table}
%------------------------------------------------------------------------------

%-------------------------------------------------------------------------------- G31.9+0.0
\begin{table}[h!]
\small
\centering
\caption{Integrated flux densities compiled for the SNR G31.9+0.0, used to trace the spectrum presented in Fig.~\ref{74-spectra}i.}
\label{flux_densities_list_G31.9}
%--------------------------------------------------------------------------------
\begin{tabular}{c l l}\hline\hline\\[-5pt]
\multicolumn{3}{c}{\bf G31.9+0.0 (3C~391)}                          \\[4pt]\hline
%--------------------------------------------------------------------------------
  Frequency   &     Scaled     flux          & \multirow{2}{*}{Reference} \\
  $[$MHz$]$   &     density     [Jy]          &                            \\\hline
    74	&	31.5	$\pm$	5.4	&	This work (from VLSSr)	\\
    74	&	28.1	$\pm$	1.8\tablefootmark{~\dag}		&	\citet{bro05-391}	\\
    80	&	26.9	$\pm$	5.0	&	\citet{caswell-71}	\\
    80	&	38.2	$\pm$	7.6	&	\citet{dulk-slee-75-to}	\\
    80	&	32.5	$\pm$	4.8	&	\citet{dulk-slee-72}	\\
    83	&	37.8	$\pm$	4.0	&	\citet{kovalenko-94}	\\
    86	&	37.2	$\pm$	2.8	&	\citet{artyukh-69}	\\
    88	&	43.7	$\pm$	8.0	&	This work (from GLEAM)	\\
   118	&	48.1	$\pm$	4.9	&	This work (from GLEAM)	\\
   155	&	44.4	$\pm$	4.1	&	This work (from GLEAM)	\\
   330	&	38.9	$\pm$	0.3\tablefootmark{~\dag}	&	\citet{bro05-391}	\\
   330	&	41.1	$\pm$	4.0	&	\citet{kassim-92}	\\
   330	&	36.1	$\pm$	0.5	&	\citet{moffett-94}	\\
   400	&	32.4	$\pm$	5.2	&	\citet{kellermann-64} \\
   408	&	30.9	$\pm$	4.6	&	\citet{kesteven-68}	\\
   408	&	33.5	$\pm$	5.0	&	\citet{green-74}	\\
   408	&	33.5	$\pm$	2.9	&	\citet{caswell-71}	\\
   408	&	33.1	$\pm$	5.0	&	\citet{kas89s} \\
   750	&	29.9	$\pm$	6.0	&	\citet{holden-caswell-69}	\\
   750	&	29.8	$\pm$	1.5	&	\citet{pauliny-toth-66}	\\
   750	&	29.8	$\pm$	3.0	&	\citet{kellermann-69}	\\
   960	&	26.3	$\pm$	2.5	&	\citet{trushkin-98}	\\
  1000	&	22.0	$\pm$	3.3\tablefootmark{~\dag}		&	\citet{milne-70}	\\
  1400	&	20.1	$\pm$	0.5	&	\citet{holden-caswell-69}	\\
  1400	&	20.8	$\pm$	0.5	&	\citet{pauliny-toth-66}	\\
  1400	&	21.4	$\pm$	1.1	&	\citet{kellermann-69}	\\
  1400	&	20.0	$\pm$	2.0\tablefootmark{~\dag}		&	\citet{goss-79}	\\
  1400	&	18.0	$\pm$	3.6	&	\citet{gardner-75} \\
  1410	&	18.8	$\pm$	2.8	&	\citet{gardner-69}	\\
  1410	&	18.5	$\pm$	3.7	&	\citet{beard-kerr-69}	\\
  1414	&	20.8	$\pm$	2.1	&	\citet{altenhoff-70}	\\
  1465	&	20.2	$\pm$	0.1\tablefootmark{~\dag}		&	\citet{bro05-391}	\\
  1468	&	22.6	$\pm$	2.3	&	\citet{moffett-94}	\\
  1667	&	16.4	$\pm$	1.5\tablefootmark{~\dag}		&	\citet{caswell-71}	\\
  2695	&	13.9	$\pm$	1.4	&	\citet{altenhoff-70}	\\
  2695	&	13.9	$\pm$	0.7	&	\citet{kellermann-69}	\\
  2700	&	14.0	$\pm$	1.4	&	\citet{becker-kundu-76}	\\
  3900	&	10.9	$\pm$	1.1	&	\citet{trushkin-98}	\\
  4800	&	 8.8	$\pm$	0.6	&	\citet{sun11}	\\
  4875	&	 9.5	$\pm$	1.4	&	\citet{downes-80}	\\
  4848	&	10.6	$\pm$	1.1	&	\citet{moffett-94}	\\
  5000	&	 9.6	$\pm$	1.9	&	\citet{reifenstein-70}	\\
  5000	&	 9.2	$\pm$	0.9	&	\citet{altenhoff-70}	\\
  5000	&	 9.7	$\pm$	0.8	&	\citet{milne-69}	\\
  5000	&	 9.8	$\pm$	1.0	&	\citet{kellermann-69}	\\
  5009	&	 9.1	$\pm$	1.4	&	\citet{shimmins-69}	\\
  6630	&	 8.6	$\pm$	0.5	&	\citet{bridle-kesteven-71}	\\
  8160	&	 7.8	$\pm$	0.1\tablefootmark{~\dag}		&	\citet{becker-kundu-76}	\\
  8800	&	 8.9	$\pm$	1.8	&	\citet{dickel-73}	\\
 10600	&	 7.4	$\pm$	0.9	&	\citet{becker-kundu-75}	\\
 10630	&	 7.1	$\pm$	0.6	&	\citet{bridle-kesteven-71}	\\
 10700	&	 7.5	$\pm$	0.8\tablefootmark{~\dag}		&	\citet{goss-79}	\\
 15500	&    5.3	$\pm$	0.5\tablefootmark{~\dag}		&	\citet{chaisson-74}	\\\hline
\end{tabular}
\tablefoot{
\tablefoottext{\dag}
{No correction factor to bring the original flux density measurement to the scale of \citet{per17} was applied because no information on the flux density calibrators was found in the original publication.}}
\end{table}
%---------------------------------------------------------------------------------

%-------------------------------------------------------------------------------- G39.2-0.3
\begin{table}[h!]
\small
\centering
\caption{Integrated flux densities compiled for the SNR G39.2$-$0.3, used to trace the spectrum presented in Fig.~\ref{74-spectra}j.}
\label{flux_densities_list_G39.2}
%--------------------------------------------------------------------------------
\begin{tabular}{c l l}\hline\hline\\[-5pt]
\multicolumn{3}{c}{\bf G39.2$-$0.3 (3C~396)}                          \\[4pt]\hline
%--------------------------------------------------------------------------------
  Frequency   &    Scaled  flux          & \multirow{2}{*}{Reference} \\
  $[$MHz$]$   &    density      [Jy]          &                            \\\hline
    25	&	28.0	$\pm$	 7.0\tablefootmark{~*}	 &	\citet{braude79}	\\
  30.9	&	45.3	$\pm$	 8.7\tablefootmark{~*}	 &	\citet{kassim-89-list}	\\
    74	&	44.8	$\pm$	 8.8 &	This work (from VLSSr)	\\
    80	&	41.1	$\pm$	 6.7 &	\citet{dulk-slee-75-to}	\\
    80	&	36.3	$\pm$	 5.1 &	\citet{slee-higgins-73}	\\
    80	&	36.4	$\pm$  10.3	&	\citet{slee77}	\\
    80	&	41.1	$\pm$	4.0	&	\citet{dulk-slee-72}	\\
   118	&	39.9	$\pm$	4.9	&	This work (from GLEAM)	\\
   155	&	33.9	$\pm$	3.4	&	This work (from GLEAM)	\\
   160	&	34.2	$\pm$	5.1	&	\citet{slee77}	\\
   160	&	37.5	$\pm$	6.9	&	\citet{dulk-slee-75-to}	\\
   200	&	28.5	$\pm$	2.8	&	This work (from GLEAM)	\\
   327	&	23.5	$\pm$	2.3	&	\citet{patnaik-90}	\\
   408	&	23.9	$\pm$	2.4	&	\citet{fanti-74}	\\
   750	&	19.6	$\pm$	3.9	&	\citet{holden-caswell-69}	\\
   750	&	17.5	$\pm$	1.0	&	\citet{pauliny-toth-66}	\\
   750	&	17.5	$\pm$	1.8	&	\citet{kellermann-69}	\\
   960	&	14.5	$\pm$	2.9	&	\citet{wilson-63}	\\
  1000	&	16.0	$\pm$	2.4\tablefootmark{~\dag}		&	\citet{milne-70}	\\
  1400	&	13.4	$\pm$	2.0	&	\citet{holden-caswell-69}	\\
  1400	&	13.6	$\pm$	3.4	&	\citet{milne-hill-69}	\\
  1400	&	16.0	$\pm$	3.2\tablefootmark{~\dag}		&	\citet{downes-71}	\\
  1400	&	14.2	$\pm$	0.7	&	\citet{kellermann-69}	\\
  1400	&	14.0	$\pm$	2.1\tablefootmark{~\dag}		&	\citet{becker-helfand-87}	\\
  1410	&	16.0	$\pm$	3.2\tablefootmark{~\dag}		&	\citet{milne-69}	\\
  1414	&	16.7	$\pm$	1.7	&	\citet{altenhoff-70}	\\
  1420	&	15.5	$\pm$	3.1	&	\citet{gardner-75}	\\
  1465	&	15.6	$\pm$	1.6	&	\citet{patnaik-90}	\\
  1635	&	14.7	$\pm$	1.5	&	\citet{patnaik-90}	\\
  1720	&	14.4	$\pm$	0.8	&	\citet{downes-81}	\\
  2300	&	13.2	$\pm$	1.0	&	\citet{trushkin-98}	\\
  2650	&	12.7	$\pm$	2.5\tablefootmark{~\dag}		&	\citet{milne-69}	\\
  2695	&	11.0	$\pm$	1.1	&	\citet{altenhoff-70}	\\
  2695	&	11.1	$\pm$	0.6	&	\citet{horton-69}	\\
  2695	&	11.4	$\pm$	1.1\tablefootmark{~\dag}		&	\citet{reich-84}	\\
  2700	&	11.9	$\pm$	1.8	&	\citet{day-70}	\\
  3240	&	11.4	$\pm$	0.7	&	\citet{hughes-butler-69}	\\
  3900	&	11.1	$\pm$	1.0	&	\citet{trushkin-98}	\\
  4800	&	 8.7	$\pm$	0.5	&	\citet{sun11}	\\
  4875	&	 9.3	$\pm$	0.9	&	\citet{altenhoff-79}	\\
  5000	&	 8.8	$\pm$	1.8	&	\citet{gardner-75}	\\
  5000	&	 8.5	$\pm$	1.7	&	\citet{milne-dickel-75}	\\
  5000	&	 8.9	$\pm$	1.8	&	\citet{reifenstein-70}	\\
  5000	&	 9.2	$\pm$	2.8	&	\citet{altenhoff-70}	\\
  5000	&	 9.8	$\pm$	0.5	&	\citet{kellermann-69}	\\
  5000	&	 8.7	$\pm$	1.3	&	\citet{dulk-slee-72}	\\
  5000	&	 9.0	$\pm$	1.4\tablefootmark{~\dag}		&	\citet{becker-helfand-87}	\\
  6630	&	 9.6	$\pm$	0.7	&	\citet{hughes-butler-69}	\\
  7700	&	 9.1	$\pm$	0.9	&	\citet{trushkin-98}	\\
  8400	&	 8.6	$\pm$	0.4	&	\citet{cru16}	\\
 10630	&	 7.6	$\pm$	0.9	&	\citet{hughes-butler-69}	\\
 13500	&	 6.3	$\pm$	0.3\tablefootmark{~\dag}		&	\citet{cru16}	\\
 33000	&	 5.2	$\pm$	0.3	&	\citet{cru16}	\\\hline
\end{tabular}
\tablefoot{
\tablefoottext{\dag}
{No correction factor to bring the original flux density measurement to the scale of \citet{per17} was applied because no information on the flux density calibrators was found in the original publication.}\\
\tablefoottext{*}{No correction to \citet{per17} was applied because the measurement is out of the 50~MHz-50~GHz frequency range where their scale is valid. }}
\end{table}
%--------------------------------------------------------------------------------

%------------------------------------------------------------------------- G41.1-0.3
\begin{table}[h!]
\small
\centering
\caption{Integrated flux densities compiled for the SNR G41.1$-$0.3, used to trace the spectrum presented in Fig.~\ref{74-spectra}k.}
\label{flux_densities_list_G41.1}
%--------------------------------------------------------------------------------
\begin{tabular}{c l l}\hline\hline\\[-5pt]
\multicolumn{3}{c}{\bf G41.1$-$0.3 (3C~397)}                          \\[4pt]\hline
%--------------------------------------------------------------------------------
  Frequency   &     Scaled    flux          & \multirow{2}{*}{Reference} \\
  $[$MHz$]$   & density         [Jy]          &                            \\\hline
  26.3	&	32.0	$\pm$	 3.0\tablefootmark{~*}	&	\citet{erickson-cronin-65}	\\
  30.9	&	40.7	$\pm$	 8.1\tablefootmark{~*}	&	\citet{kassim-88}	\\
    74	&	68.9	$\pm$	10.6	&	This work (from VLSSr)	\\
    80	&	64.2	$\pm$	 9.0	&	\citet{slee-higgins-75}	\\
    80	&	60.8	$\pm$	 6.0	&	\citet{dulk-slee-75-to}	\\
  81.5	&	73.8	$\pm$	11.1	&	\citet{readhead-hewish-74}	\\
    83	&	57.7	$\pm$	11.9	&	\citet{kovalenko-94}	\\
    86	&	66.6	$\pm$	 3.2	&	\citet{artyukh-69}	\\
    88	&	57.0	$\pm$	 7.0	&	This work (from GLEAM)	\\
   102	&	59.2	$\pm$	17.8	&	\citet{kovalenko-94}	\\
   111	&	70.9	$\pm$	11.8	&	\citet{kovalenko-94}	\\
   160	&	52.3	$\pm$	 3.7	&	\citet{slee77}	\\
   178	&	61.5	$\pm$	12.3	&	\citet{kellermann-69}	\\
   610	&	36.9	$\pm$	 7.4	&	\citet{holden-caswell-69}	\\
   612	&	38.7	$\pm$	 3.9	&	\citet{conway-65}	\\
  1400	&	28.2	$\pm$	 4.2	&	\citet{holden-caswell-69}	\\
  1400	&	30.1	$\pm$	 6.0	&	\citet{kellermann-69}	\\
  1414	&	25.0	$\pm$	 2.5	&	\citet{altenhoff-70}	\\
  2695	&	20.0	$\pm$	20.0	&	\citet{kellermann-69}	\\
  2700	&	20.3	$\pm$	 3.7	&	\citet{willis-73}	\\
  2700	&	21.2	$\pm$	 1.2	&	\citet{velusamy-kundu-74}	\\
  2700	&	19.9	$\pm$	 3.0	&	\citet{day-70}	\\
  4800	&	18.3	$\pm$	 1.1	&	\citet{sun11}	\\
  5000	&	15.4	$\pm$	 1.5	&	\citet{altenhoff-70}	\\
  5000	&	14.8	$\pm$	 3.0	&	\citet{kellermann-69}	\\
  5000	&	18.9	$\pm$	 3.8	&	\citet{kundu-74}	\\
 10600	&	10.9	$\pm$	 1.8	&	\citet{dickel-denoyer-75}	\\
 10700	&	13.0	$\pm$	 2.6	&	\citet{kundu-74}	\\\hline
\end{tabular}
\tablefoot{\tablefoottext{*}{No correction to \citet{per17} was applied because the measurement is out of the 50~MHz-50~GHz frequency range where their scale is valid. }}
\end{table}
%-------------------------------------------------------------------------

%-------------------------------------------------------------------------- G43.3-0.2
\begin{table}[h!]
\small
\caption{Integrated flux densities compiled for the SNR G43.3$-$0.2, used to trace the spectrum presented in Fig.~\ref{74-spectra}l.}
\label{flux_densities_list_G43.3}
%--------------------------------------------------------------------------------
\begin{tabular}{c l l}\hline\hline\\[-5pt]
\multicolumn{3}{c}{\bf G43.3$-$0.2 (W49B)}                          \\[4pt]\hline
%--------------------------------------------------------------------------------
  Frequency   &    Scaled flux          & \multirow{2}{*}{Reference} \\
  $[$MHz$]$   & density         [Jy]          &                            \\\hline
   38	&	16.0	$\pm$	2.4\tablefootmark{~*}	    &	\citet{holden-caswell-69}	\\
    74	&	64.0	$\pm$  10.1	&	This work (from VLSSr)	\\
    80	&	67.0	$\pm$  12.0	&	\citet{dickel-73}	\\
  81.5	&	74.3	$\pm$  11.1	&	\citet{readhead-hewish-74}	\\
    86	&	66.3	$\pm$	3.2	    &	\citet{artyukh-69}	\\
    88	&	36.8	$\pm$	8.0	    &	This work (from GLEAM)	\\
   118	&	69.7	$\pm$	5.5	    &	This work (from GLEAM)	\\
   160	&	68.1	$\pm$	5.9	    &	\citet{dulk-slee-75-to}	\\
   160	&	67.4	$\pm$  10.0	&	\citet{slee77}	\\
   178	&	67.4	$\pm$	3.4	    &	\citet{kellermann-69}	\\
   400	&	50.0	$\pm$  10.0\tablefootmark{~\dag}		&	\citet{downes-71}	\\
   408	&	51.6	$\pm$	7.7 	&	\citet{shaver-69}	\\
   960	&	37.4	$\pm$	3.5	    &	\citet{trushkin-98}	\\
  1000	&	33.0	$\pm$	5.0\tablefootmark{~\dag}		&	\citet{milne-70}	\\
  1400	&	30.0	$\pm$	6.0\tablefootmark{~\dag}		&	\citet{downes-71}	\\
  1414	&	30.2	$\pm$	3.0	&	\citet{altenhoff-70}	\\
  2650	&	25.3	$\pm$	5.1	&	\citet{milne-hill-69}	\\
  2695	&	19.9	$\pm$	2.0	&	\citet{altenhoff-70}	\\
  3250	&	21.8	$\pm$	3.1	&	\citet{hughes-butler-69}	\\
  3900	&	18.2	$\pm$	1.9	&	\citet{trushkin-98}	\\
  4800	&	18.8	$\pm$	1.0	&	\citet{sun11}	\\
  4850	&	16.2	$\pm$	0.0\tablefootmark{~\dag}		&	\citet{taylor-92}	\\
  4875	&	17.9	$\pm$	2.7	&	\citet{downes-80}	\\
  5000	&	14.3	$\pm$	1.4	&	\citet{kellermann-69}	\\
  5000	&	14.3	$\pm$	1.4	&	\citet{altenhoff-70}	\\
  5000	&	15.5	$\pm$	3.1	&	\citet{goss-shaver-70}	\\
  5000	&	16.8	$\pm$	3.4	&	\citet{shaver-goss-70b}	\\
  5000	&	20.0	$\pm$	4.0\tablefootmark{~\dag}		&	\citet{downes-71}	\\
 10630	&	13.1	$\pm$	2.0	&	\citet{hughes-butler-69}	\\
 11200	&	10.0	$\pm$	1.4	&	\citet{trushkin-98}	\\
 32000	&	 6.2	$\pm$	0.3	&	\citet{morsi-reich-87}	\\\hline
\end{tabular}
\tablefoot{
\tablefoottext{\dag}
{No correction factor to bring the original flux density measurement to the scale of \citet{per17} was applied because no information on the flux density calibrators was found in the original publication.}\\
\tablefoottext{*}{No correction to \citet{per17} was applied because the measurement is out of the 50~MHz-50~GHz frequency range where their scale is valid. }}
\end{table}
%--------------------------------------------------------------------------

%------------------------------------------------------------------------- G120.1+1.4
\begin{table}[h!]
\small
\caption{Integrated flux densities compiled for the SNR G120.1+1.4, used to trace the spectrum presented in Fig.~\ref{74-spectra}m.}
\label{flux_densities_list_G120.1}
%-------------------------------------------------------------------------
\begin{tabular}{c l l}\hline\hline\\[-5pt]
\multicolumn{3}{c}{\bf G120.1+1.4 (Tycho)}                        \\[4pt]\hline
%-------------------------------------------------------------------------
 Frequency   &   Scaled flux         &  \multirow{2}{*}{Reference} \\
  $[$MHz$]$  &       density  [Jy]         &                             \\\hline
  14.7	&	770.0	$\pm$	131.0\tablefootmark{~*}	&	\citet{viny87}	\\
  16.7	&	680.0	$\pm$	88.0\tablefootmark{~*}	&	\citet{viny87}	\\
    20	&	600.0	$\pm$	78.0\tablefootmark{~*}	&	\citet{viny87}	\\
 22.25	&	590.0	$\pm$	47.0\tablefootmark{~*}	&	\citet{roger-69}	\\
    25	&	515.0	$\pm$	62.0\tablefootmark{~*}	&	\citet{viny87}	\\
  48.3	&	334.0	$\pm$	33.0\tablefootmark{~*}	&	\citet{arias+19}	\\
    67	&	275.0	$\pm$	27.0\tablefootmark{~\dag}	&	\citet{arias+19}	\\
    74	&	255.0	$\pm$	38.8	&	This work (from VLSSr)	\\
 102.5	&	227.6	$\pm$	16.4\tablefootmark{~\dag}	&	\citet{viny87}	\\
 144.6	&	163.0	$\pm$	16.0\tablefootmark{~\dag}	&	\citet{arias+19}	\\
   178	&	143.8	$\pm$	 7.2	&	\citet{kellermann-69}	\\
   178	&	135.8	$\pm$	20.4	&	\citet{bennett-63}	\\
   232	&	135.5	$\pm$	20.3	&	\citet{zhang-97}	\\
   327	&	101.2	$\pm$	10.1	&	\citet{arias+19}	\\
   408	&	100.8	$\pm$	15.1	&	\citet{fanti-74}	\\
   408	&	86.0	$\pm$	 5.0\tablefootmark{~\dag}	&	\citet{kothes-06}	\\
   408	&	83.7	$\pm$	16.7	&	\citet{green-74}	\\
   612	&	66.8	$\pm$	 3.4	&	\citet{conway-65}	\\
   750	&	63.8	$\pm$	 3.2	&	\citet{kellermann-69}	\\
   960	&	55.5	$\pm$	 0.8	&	\citet{trushkin-98}	\\
   960	&	54.5	$\pm$	 0.8	&	\citet{trushkin-87}	\\
  1000	&	58.0	$\pm$	 8.7\tablefootmark{~\dag}	&	\citet{milne-70}	\\
  1382	&	41.7	$\pm$	 4.2\tablefootmark{~\dag}	&	\citet{arias+19}	\\
  1400	&	43.8	$\pm$	 2.1	&	\citet{conway-65}	\\
  1400	&	44.4	$\pm$	 2.2	&	\citet{kellermann-69}	\\
  1410	&	45.1	$\pm$	 4.5	&	\citet{reich-97}	\\
  1420	&	40.5	$\pm$	 1.5\tablefootmark{~\dag}	&	\citet{kothes-06}	\\
  2300	&	29.2	$\pm$	 0.8	&	\citet{trushkin-98}	\\
  2300	&	28.7	$\pm$	 0.8	&	\citet{trushkin-87}	\\
  2695	&	29.5	$\pm$	 1.5	&	\citet{kellermann-69}	\\
  2695	&	26.2	$\pm$	 1.8	&	\citet{horton-69}	\\
  3650	&	22.9	$\pm$	 0.3	&	\citet{trushkin-87}	\\
  3900	&	22.4	$\pm$	 0.3	&	\citet{trushkin-87}	\\
  3900	&	20.0	$\pm$	 0.5	&	\citet{trushkin-98}	\\
  4800	&	19.7	$\pm$	 2.0	&	\citet{sun11}	\\
  4995	&	16.9	$\pm$	 0.9	&	\citet{horton-69}	\\
  5000	&	20.7	$\pm$	 1.0	&	\citet{kellermann-69}	\\
  5000	&	19.7	$\pm$	 1.0	&	\citet{reich-14}	\\
  7700	&	14.6	$\pm$	 0.5	&	\citet{trushkin-98}	\\
 10700	&	12.8	$\pm$	 0.8	&	\citet{klein-79}	\\
 11200	&	11.7	$\pm$	 0.5	&	\citet{trushkin-98}	\\
 15000	&	10.7	$\pm$	 1.0	&	\citet{klein-79}	\\
 15000	&	10.6	$\pm$	 0.3	&	\citet{hurley-walker-09}	\\
 15700	&	10.1	$\pm$	 0.3	&	\citet{hurley-walker-09}	\\
 16400	&	 9.1	$\pm$	 0.2	&	\citet{hurley-walker-09}	\\
 17100	&	 8.0	$\pm$	 0.2	&	\citet{hurley-walker-09}	\\
 21400	&	 8.8	$\pm$	 0.9\tablefootmark{~\dag}	&	\citet{lor19}	\\
 44000	&	 5.2	$\pm$	 0.3\tablefootmark{~\dag}	&	\citet{planck2016}	\\
 70000	&	 4.4	$\pm$	 0.3\tablefootmark{~*}	&	\citet{planck2016}	\\\hline
\end{tabular}
\tablefoot{
\tablefoottext{\dag}
{No correction factor to bring the original flux density measurement to the scale of \citet{per17} was applied because no information on the flux density calibrators was found in the original publication.}\\
\tablefoottext{*}{No correction to \citet{per17} was applied because the measurement is out of the 50~MHz-50~GHz frequency range where their scale is valid. }}
\end{table}
%-------------------------------------------------------------------------

%------------------------------------------------------------------------- G130.7+3.1
\begin{table}[h!]
\small
\caption{Integrated flux densities compiled for the SNR G130.7+3.1, used to trace the spectrum presented in Fig.~\ref{74-spectra}n.}
\label{flux_densities_list_G130.7}
%-------------------------------------------------------------------------
\begin{tabular}{c l l}\hline\hline\\[-5pt]
\multicolumn{3}{c}{\bf G130.7+3.1 (3C~58)}                         \\[4pt]\hline
%-------------------------------------------------------------------------
 Frequency   &      Scaled flux         &  \multirow{2}{*}{Reference} \\
  $[$MHz$]$  &       density  [Jy]         &                             \\\hline
     38	&	37.9	$\pm$	8.0\tablefootmark{~*}	&	\citet{rees90}	\\
     74	&	34.6	$\pm$	5.3	&	This work (from VLSSr)	\\
     74	&	33.6	$\pm$	6.7	&	\citet{bietenholz-01}	\\
     83	&	40.8	$\pm$	5.0	&	\citet{kovalenko-94}	\\
     86	&	37.2	$\pm$	1.8 &	\citet{artyukh-69}	\\     
    102	&	36.5	$\pm$	4.9	&	\citet{kovalenko-94}	\\
    111	&	35.4	$\pm$	4.9	&	\citet{kovalenko-94}	\\
    151	&	32.0	$\pm$	3.6	&	\citet{green-86}	\\
    327	&	33.9	$\pm$	5.1	&	\citet{bietenholz-01}	\\
    400	&	33.0	$\pm$	6.6\tablefootmark{~\dag}	&	\citet{downes-71}	\\
    400	&	33.8	$\pm$	2.1	&	\citet{kellermann-64}	\\
    408	&	32.1	$\pm$	6.4	&	\citet{green-74}	\\
    408	&	36.1	$\pm$	4.3	&	\citet{kassim-89-list}	\\
    408	&	35.8	$\pm$	5.4	&	\citet{fanti-74}	\\
    408	&	32.2	$\pm$	2.0\tablefootmark{~\dag}	&	\citet{kothes-06}	\\
    750	&	35.4	$\pm$	1.8	&	\citet{kellermann-69}	\\
    750	&	35.2	$\pm$	0.4	&	\citet{pauliny-toth-66}	\\
    958	&	33.5	$\pm$	3.3	&	\citet{conway-63}	\\
    960	&	31.8	$\pm$	6.4	&	\citet{wilson-63}	\\
    960	&	32.1	$\pm$	1.4	&	\citet{harris-roberts-60}	\\
    960	&	33.7	$\pm$	1.5	&	\citet{trushkin-98}	\\
   1400	&	34.5	$\pm$	1.7	&	\citet{kellermann-69}	\\
   1400	&	33.0	$\pm$	6.6\tablefootmark{~\dag}	&	\citet{downes-71}	\\
   1400	&	34.5	$\pm$	0.5	&	\citet{pauliny-toth-66}	\\
   1420	&	31.9	$\pm$	1.0\tablefootmark{~\dag}	&	\citet{kothes-06}	\\
   1420	&	34.2	$\pm$	5.1\tablefootmark{~\dag}	&	\citet{weiler-seielstad-71}	\\
   1420	&	29.3	$\pm$	5.9	&	\citet{galt-68}	\\
   1446	&	33.0	$\pm$	5.0\tablefootmark{~\dag}	&	\citet{reynolds-aller-85}	\\
   2300	&	31.9	$\pm$	2.0	&	\citet{trushkin-98}	\\
   2695	&	31.3	$\pm$	1.6	&	\citet{kellermann-69}	\\
   2695	&	30.8	$\pm$	3.1	&	\citet{green-86}	\\
   2695	&	26.9	$\pm$	1.7	&	\citet{horton-69}	\\
   2880	&	30.2	$\pm$	4.5\tablefootmark{~\dag}	&	\citet{weiler-seielstad-71}	\\
   3200	&	30.1	$\pm$	3.0	&	\citet{conway-63}	\\
   3650	&	31.7	$\pm$	1.0	&	\citet{trushkin-98}	\\
   3900	&	30.2	$\pm$	1.0	&	\citet{trushkin-98}	\\
   4800	&	31.2	$\pm$	3.0	&	\citet{sun11}	\\
   4995	&	28.1	$\pm$	1.6	&	\citet{horton-69}	\\
   5000	&	26.4	$\pm$	1.3	&	\citet{kellermann-69}	\\
   5000	&	27.0	$\pm$	5.4\tablefootmark{~\dag}	&	\citet{downes-71}	\\
   5000	&	26.4	$\pm$	1.0	&	\citet{pauliny-toth-66}	\\
  14200	&	25.5	$\pm$	1.2	&	\citet{hurley-walker-09}	\\
  15000	&	26.7	$\pm$	0.5\tablefootmark{~\dag}	&	\citet{green-75}	\\
  15000	&	24.9	$\pm$	1.2	&	\citet{hurley-walker-09}	\\
  15700	&	24.1	$\pm$	1.1	&	\citet{hurley-walker-09}	\\
  16400	&	22.5	$\pm$	1.0	&	\citet{hurley-walker-09}	\\
  17100	&	21.9	$\pm$	1.0	&	\citet{hurley-walker-09}	\\
  17900	&	23.1	$\pm$	1.1	&	\citet{hurley-walker-09}	\\
  30000	&	22.2	$\pm$	2.2\tablefootmark{~\dag}	&	\cite{planck2016}	\\
  32000	&	21.6	$\pm$	1.3	&	\citet{morsi-reich-87}	\\
  44000	&	16.4	$\pm$	1.6\tablefootmark{~\dag}	&	\cite{planck2016}	\\
  70000	&	14.2	$\pm$	1.4\tablefootmark{~*}	&	\cite{planck2016}	\\
  84200	&	15.0	$\pm$	2.0\tablefootmark{~*}	&	\citet{salter-89}	\\
 100000	&	12.7	$\pm$	1.3\tablefootmark{~*}	&	\cite{planck2016}	\\
 143000	&	10.8	$\pm$	1.1\tablefootmark{~*}	&	\cite{planck2016}	\\
 217000	&	 8.4	$\pm$	0.8\tablefootmark{~*}	&	\cite{planck2016}	\\\hline
\end{tabular}
\tablefoot{
\tablefoottext{\dag}
{No correction factor to bring the original flux density measurement to the scale of \citet{per17} was applied because no information on the flux density calibrators was found in the original publication.}\\
\tablefoottext{*}{No correction to \citet{per17} was applied because the measurement is out of the 50~MHz-50~GHz frequency range where their scale is valid. }}
\end{table}
%-------------------------------------------------------------------------

\end{document}